\documentclass[pdflatex,sn-mathphys]{sn-jnl}

\newlength{\figurewidth}
\jyear{2022}%

\theoremstyle{thmstyleone}%
\theoremstyle{thmstyletwo}%

\theoremstyle{thmstylethree}%

\begin{document}

\title[Article Title]{Crystal Chemistry at High Pressure}


\author[1]{\fnm{Katerina P.} \sur{Hilleke}} \email{khilleke@gmail.com}

\author*[1]{\fnm{Eva} \sur{Zurek}}\email{ezurek@buffalo.edu}

\affil[1]{\orgdiv{Department of Chemistry}, \orgname{State University of New York at Buffalo}, \orgaddress{\street{777 Natural Sciences Complex}, \city{Buffalo}, \postcode{14260-3000}, \state{NY}, \country{USA}}}

\maketitle

\section{Introduction}\label{sec1}

Jupiter, the largest planet in our solar system, is a gas giant comprised mostly of hydrogen and helium. After penetrating its atmosphere, a mixture of hydrogen and helium, methane and ammonia, one would find a massive sea of hydrogen -- behaving very differently than one might expect from general chemistry classes~\cite{Guillot:2005}. The hydrogen layer surrounding Jupiter's core, under immense pressure and at high temperature, is metallic. This is just one example of how the pressure variable is so important in determining chemical and physical behavior, bearing consequences for the evolution and dynamics of planetary interiors \cite{Zaghoo:2017a}.

\begin{figure}[!h]
\begin{center}
\includegraphics[width=\figurewidth]{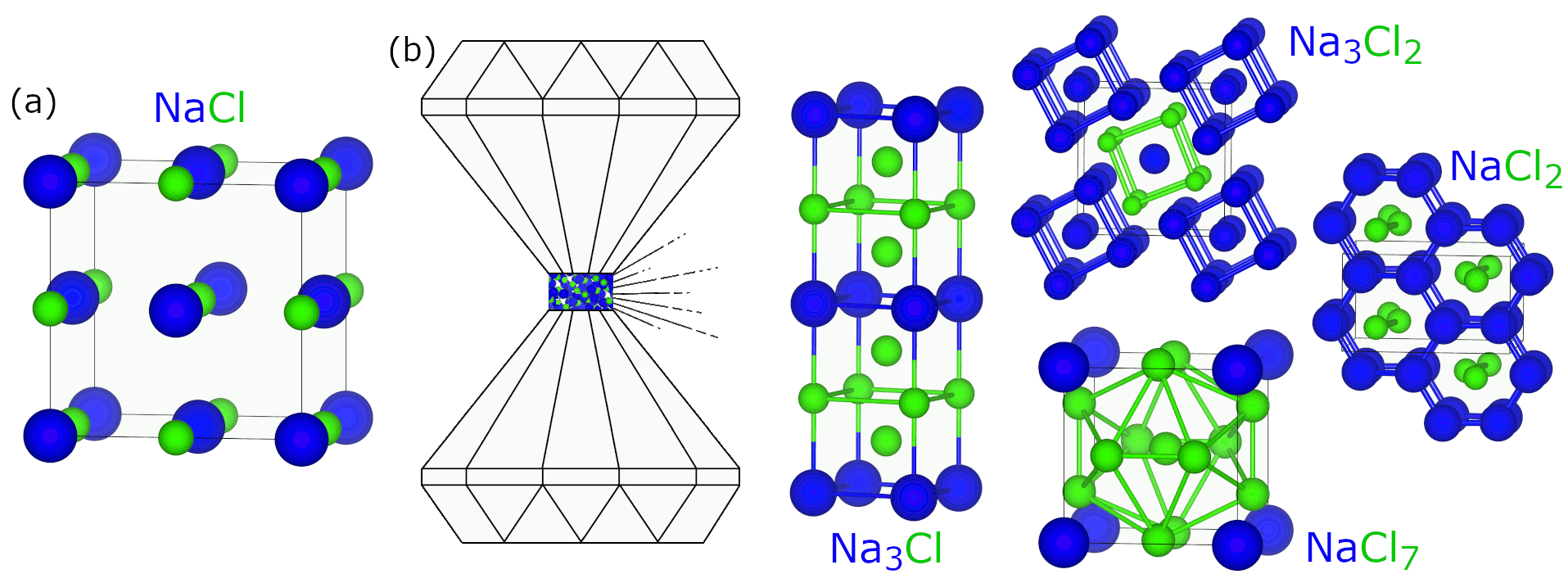}
\end{center}
\caption{The familiar crystal structure of NaCl, table salt, (a) is the only known stable compound of sodium and chlorine at ambient pressure. Under pressures exerted by a diamond anvil cell (b), a variety of additional stoichiometries and crystal structures decorate the sodium-chlorine phase diagram, including the $P4/mmm$ Na$_3$Cl, $P4/m$ Na$_3$Cl$_2$, $Imma$ NaCl$_2$, and $Pm\bar{3}$ NaCl$_7$~\cite{Dong:2017} structures shown here.
\label{fig:intro}}
\end{figure}

Under pressure, the behavior of the elements as well as the compounds they form can change drastically from what we, as beings living and learning at 1~atmosphere, are used to. Hydrogen, as described above, can become metallic at high pressure, joining its Group I cousins the alkali metals. It may be perplexing then to learn that the alkali metals lithium~\cite{Matsuoka:2009} and sodium~\cite{Ma:2009,Polsin:2022} become semiconducting and insulating, respectively, when squeezed. Potassium starts to behave more like a transition metal~\cite{Bukowinski:1976,Atou:1996,Parker:1996}. When combined with one another, the unexpected behavior of the elements under pressure makes for correspondingly curious compounds. Sodium chlorides with vastly divergent stoichiometries from the typical 1:1 (Figure~\ref{fig:intro}) have been predicted and synthesized~\cite{Zhang:2013}, stubbornly inert helium atoms form a compound with sodium~\cite{Dong:2017}, and metal superhydrides wherein the hydrogen atoms coalesce into clathrate-like networks have been reported~\cite{geballe2018synthesis}. Various ``rules" for the behavior of materials at high pressure have been proposed by Prewitt and Downs~\cite{Prewitt:1998}, Grochala and co-workers~\cite{Grochala:2007a}, and Zhang \emph{et. al.}~\cite{Zhang:2017}, to name a few -- and have undergone progressive revision as more is uncovered about the structures and materials far below us in the Earth's core and far away in the planets and stars. 

Over the past century, experimental methods have evolved to create progressively higher pressures in a laboratory setting, allowing us to directly probe the behavior of materials under extreme conditions. Pressures in the megabar range can now be routinely achieved, albeit still requiring delicate setups~\cite{Jayaraman:1983,Bassett:2009,Li:2018c}. Diamond anvil cells (DACs) combine the superior hardness of this desired polymorph of carbon with its optical transparency, enabling the creation of the highest static pressures and interrogation of the sample via visual and spectroscopic means. Engineering advances from multistage compression apparatuses~\cite{Dubrovinsky:2012,Dubrovinskaia:2016} to toroidal DACs~\cite{Dewaele:2018,Jenei:2018} have driven the experimental ceiling for static pressures ever higher. Dynamic compression experiments under ramp or shock conditions, driven by gas guns, lases pulses, or magnetic fields can reach well into the terapascal regime~\cite{Duffy:2019,Hansen:2021,Fratanduono:2021}, allowing us to study the behavior of diamond at 5~TPa~\cite{Smith:2014} and iron at conditions thought to be in super-Earths \cite{Kraus:2022a}. Diagnostic techniques for characterizing the resulting substances can be difficult to implement in both dynamic and static compression, often requiring theoretical support for their interpretation~\cite{Zurek:2014i}.

Far cheaper than high-pressure experiments are computations modeling high-pressure systems. Band structures, phonons, Raman or infra-red spectra and numerous material properties can all be calculated without so much as stepping foot into a laboratory. Crystal structure prediction (CSP) techniques, not weighted down by preconceived, atmospheric pressure-based, notions of how atoms might appropriately arrange themselves in a unit cell, can identify which structures can exist under high-pressure conditions~\cite{Pickard:2009,Oganov:2011,Wang:2014}. The computational exploration of potential energy landscapes will not be discussed in our contribution; to learn about methods that can be employed to identify the global, as well as important local, minima we point the reader to an excellent chapter in this book ``Crystal Structure Prediction'' by Andreas Hermann, Lewis J.\ Conway and Chris J.\ Pickard. Exploratory calculations highlight promising phases for further experimental investigation, but computation can just as well follow experimental results, elucidating behavior and filling in the gaps. The resulting feedback loop of experiment and theory has driven the discovery and characterization of a plethora of phases ranging from the superhard~\cite{Oganov:2009,Zhang:2015b,Avery:2019} to the superconducting~\cite{Liu:2017-La-Y, Peng:2017,Kruglov:2018,geballe2018synthesis,drozdov2019superconductivity,Semenok:2021b}. 

In the following sections we build a framework for understanding the behavior of materials at high pressure, starting from the effects of pressure on the atoms themselves, driving electronic transitions and altering periodic trends. From there, the various manifestations of high pressure in the solid state are sorted into categories (which are not mutually exclusive, but rather illustrative), starting with exotic electronic structures and electrides. We discuss compounds of the noble gases and those containing elements that are immiscible at ambient pressure, as well as crystal lattices that contain bizarre geometrical motifs and bonding configurations. Finally, we survey the effects of high pressure on superconductivity, a field that has recently undergone a veritable explosion as high pressure phases toe the line of room-temperature superconductivity~\cite{Zurek:2020k}.

\section{The atom under pressure}\label{sec3}

Chemistry describes the interactions between the 118 distinct elements that are organized in the periodic table, proposed by Mendeleev while classifying elements according to their chemical properties observed at atmospheric conditions. The trends found within the periodic table can be used to compare atoms according to their size, the number of electrons surrounding their nuclei, and to make predictions as to whether those electrons are held tightly or loosely. Moreover, the periodic table allows students and researchers to predict how different elements will interact with one another: will they form compounds, emulsions or alloys,  or will they be unreactive? If reactivity is suspected, the periodic table can be used to guess if the bonds are covalent, ionic, or (usually) somewhere in between. Across the periodic table, trends in properties such as atomic radii, electronegativity, and oxidation state can be mapped, leading one to conclude that fluorine, the most electronegative element, will gain electrons in a binary phase thereby achieving an F$^-$ configuration with a filled valence shell. On the other hand, cesium, as the least electronegative element (neglecting francium, whose miniscule half-life renders it basically experimentally irrelevant), typically assumes an oxidation state of 0 or +1.\footnote{When dissolved in amines or ethers, Na, K, Rb, and Cs can assume the unusual oxidation state of -1.~\cite{Dye:2009} These so-called \emph{alkalides} are thought to form ion-pairs with solvated metal cations.~\cite{Zurek:2011c,Zurek:2020h,Abella:2021}} Yet at high pressure, several Li$_n$Cs phases have been predicted~\cite{Botana:2014} where Cs attains unusual formal oxidation states thought to be in excess of -2 due to substantial electron transfer from lithium to cesium. How can the stability of these unintuitive stoichiometries and their resulting electronic structures be rationalized?

Let us consider electronegativity a little more. Although Pauling's~\cite{Pauling:1932} is the most widely adopted, a number of metrics have been used to produce scales of electronegativity for the elements. In Pauling's formulation, electronegativity differences between pairs of atoms A and B are calculated from the homo- and heteronuclear bond dissociation energies, then referenced against the electronegativity of H being set to 2.1 (later adjusted to 2.2). In this regime, the electronegativities of fluorine, lithium, and cesium are 3.98, 0.98, and 0.79, respectively. Several other scales have been proposed, including Mulliken's ``absolute electronegativity"~\cite{Pearson:1985} being the average of the first ionization energy and electron affinity of an atom~\cite{Mulliken:1934}. Dong \emph{et al.} modify the Mulliken definition for elements under high pressure taking as a reference the homogeneous electron gas rather than the vacuum~\cite{Dong:2022}. Allen's electronegativities are derived from the average energies of the valence electrons in the atom~\cite{Allen:1989}, and a closely related scheme has recently been proposed by Rahm et al~\cite{Rahm:2019,Rahm:2021,Racioppi:2021}, where electronegativity is calculated as the average of the electron binding energies of ground state valence electrons -- approximately translated to the Allen scale if averaging over valence electrons alone. Broadly speaking, the common factor of importance in all of these is the collection of atomic orbital energies and the differences between them. Under pressure, those change. 

\begin{figure}[!h]
\begin{center}
\includegraphics[width=\figurewidth]{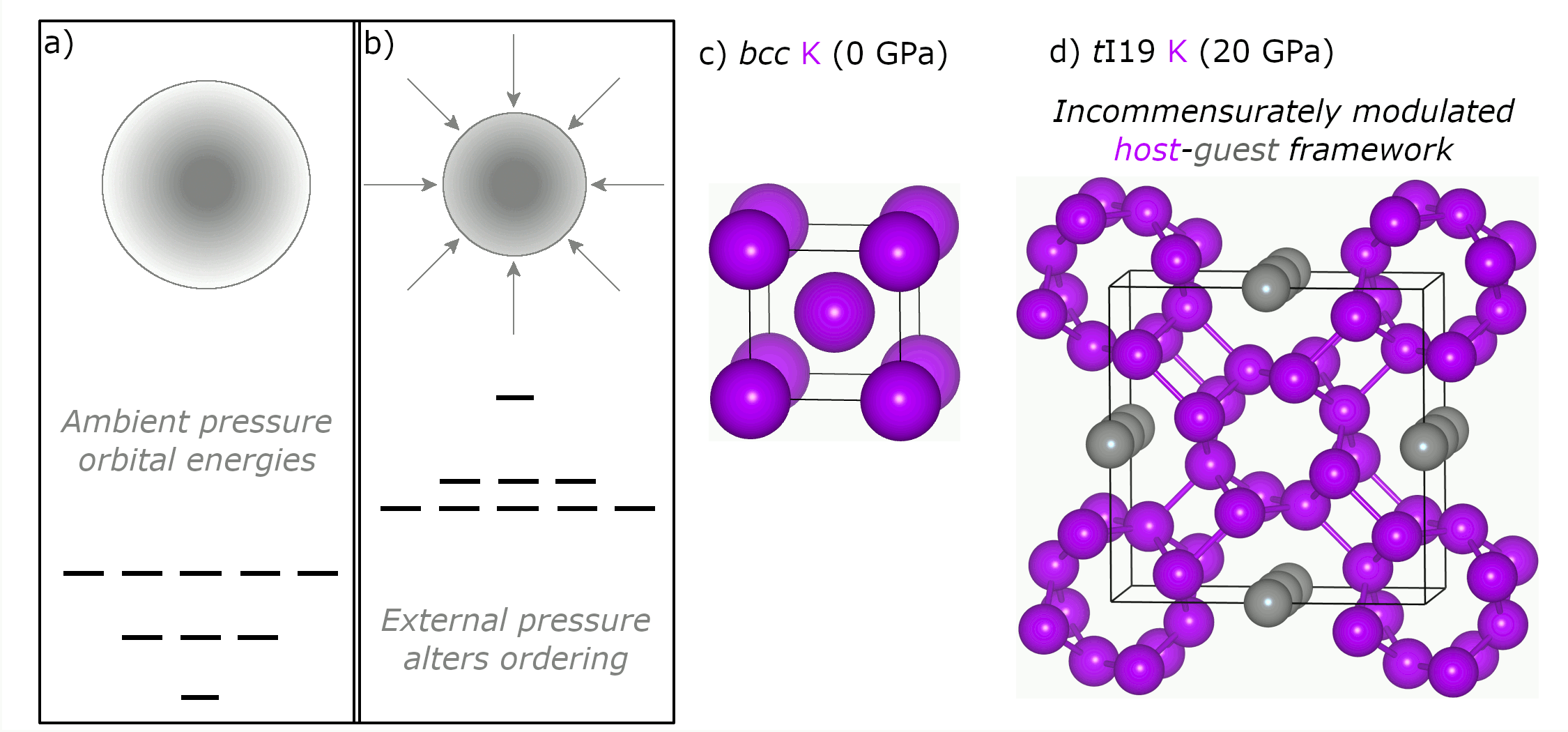}
\end{center}
\caption{External pressure on an atom raises the energy of atomic orbitals as they are constrained to a smaller space, but the rate of this increase differs between s, p, and d orbitals (a), (b), favoring s $\rightarrow$ d transitions for many of the alkali metals, including potassium. Potassium adopts the $bcc$ structure at 0~GPa (c) but at higher pressures takes on a series of complex structures including the incommensurately modulated host-guest \emph{t}I19 phase~\cite{McMahon:2006b} (d).
\label{fig:atompressure}}
\end{figure}

The prototypical model for understanding the quantized behavior of energy levels in a confined system is the particle in a box. The resulting energy levels for a particle of mass $m$ in a box of width $L$ are given by $E_n = \frac{\hbar^2n^2\pi^2}{2mL^2}$, where $n$ is the principal quantum number. To consider the effects of pressure on this model, we could simply make the box smaller by reducing the width $L$, which has the effect of increasing the $E_n$. Thus, energy levels (orbital energies) will increase under pressure. The complicating factor is that the rate of this increase is not the same for each orbital because the number of radial nodes plays a role. The peak density of a 4s orbital is further from the nucleus than a 4p orbital, since the electrons occupying the 4s orbital must maintain orthogonality to those in the 1s, 2s, and 3s orbitals, while those in the 4p orbital must only contend with the 3p and 2p orbitals (this analogy may be extended to the 4d and 4f shells). With more electron density being further from the nucleus, the electrons in the 4s orbital will feel the effects of pressure more strongly than those in the 4p orbital. This is illustrated schematically in Figure~\ref{fig:atompressure}a and \ref{fig:atompressure}b. At certain levels of confinement, electronic $n\textrm{s} \rightarrow n\textrm{p}$,  $n\textrm{s} \xrightarrow{} (n-1)\textrm{d}$, and $(n-1)\textrm{d} \xrightarrow{} (n-2)\textrm{f}$ transitions can become favorable.  This is the reason why pressure drives rearrangements of the orbital energies of an atom, with ensuing electronic transitions. (In section~\ref{subsec6}, we will explore the opportunities presented by another sort of destination orbital, this time one not centered on an atom.)

In compressed lithium, the ground state electronic configuration transitions from $1\textrm{s}^2 2\textrm{s}^1 \rightarrow 1\textrm{s}^2 2\textrm{p}^1$, while cesium undergoes a $6\textrm{s}^1 \rightarrow 5\textrm{d}^1$ pressure induced transformation. The exact pressures at which these electronic changes occur depend, of course, on the chemical environment of the lithium or cesium atoms -- and in calculations, on the theoretical methodology. In Cs, the $\textrm{s}\rightarrow\textrm{d}$ transition is thought to drive the transformation to the complex Cs-III structure with 84 atoms in the unit cell that is stable near 4.2~GPa ~\cite{Sternheimer:1950,Louie:1974,Takemura:1982,Connerade:2000,Rahm:2019b,Zurek:2005a}, while in lithium the pressure at which the s-p mixing occurs is believed to be somewhat higher, in the megabar to multimegabar range, dependent on the chemical environment and level of theory~\cite{Neaton:1999,Hanfland:2000,Naumov:2015,Rahm:2019b}. Another example includes potassium, whose complex phase diagram under pressure (see, \emph{e.g.}\ Figure~\ref{fig:atompressure}c,d),  has been in part attributed to this s~$\rightarrow$~d electronic transition.~\cite{Degtyareva:2014} The energies of core orbitals can also become relevant; for example, in Cs-VI the 5p bands hybridize with the 6s \cite{Takemura:2000,McMahan:1984}, making them accessible for chemical interactions, and the core orbitals of K have also been proposed to be key to its structural diversity.~\cite{Degtyareva:2014} Overall, one effect of pressure on the alkali metals is that their electronegativities undergo quite a rearrangement, ending up with cesium being more electronegative than lithium~\cite{Rahm:2019b,Dong:2022}. From this perspective, the formation of cesium anions in the Li$_n$Cs phases begins to make sense.

Among the $d$-block similar reorderings are predicted to take place, with the group 10 metals Ni, Pd, and Pt preferring $d^{10}$ closed-shell configurations, as compared to the $s^2d^8$ favored at ambient conditions, while the group 11 and 12 metals become electron donors~\cite{Dong:2022}. The former transition is associated with a spike in the estimated chemical hardness -- obtained as half of the HOMO-LUMO gap -- of Ni, Pd, and Pt, reaching values comparable to some of the noble gases. This is contrary to the general trend where the hardness of most of the elements in the periodic table decreases with pressure as energy levels become closer to one another~\cite{Dong:2022}. The resulting changes in the relative hardness of pairs of elements can lead to changes in compound stability arising from HSAB (hard-soft acid-base) arguments, and the appearance of strange multicenter bonding manifolds in certain high pressure phases have been linked to general increases in softness~\cite{Hooper:2012}. 

From transition-metal-like behavior in the \emph{s}-block to relative inertness in the formerly-$d^8$-transition metals, atoms under compression can behave very differently from their ambient-pressure selves, and the consequences for materials under pressure are far-reaching. We will now explore some of the wild and wonderful structures and phenomena that result.

\section{The crystal under pressure}\label{sec4}

At ambient pressure a majority of the solid, metallic elements of the periodic table adopt very simple, symmetric, structures that are close-packed. The most stable geometries are those that minimize the free energy. However, at low temperatures the entropic contributions between different solid phases are typically negligible, so computational studies often employ the enthalpy to determine the structures that are preferred. With increasing pressure the enthalpy, consisting of the internal energy and pressure-volume terms ($H=U+PV$), becomes dominated by the $PV$ term. It would be natural, therefore, to imagine that close-packed structures with increased coordination numbers become preferred at high pressures. The reality, however, does not coincide with our expectations. For example, within cesium the nearest neighbor coordination number first increases from 8 (bcc) to 12 (fcc), then decreases to about 10 (Cs-III), 8 (Cs-IV) and finally increases again to 10/11 (Cs-V) and 12 (Cs-VI). These structural transitions are believed to be driven by the previously discussed pressure-induced $s\rightarrow d$ valence electronic transition within the constituent atoms, which causes the interatomic distances to become smaller compared to the ranges of the wavefunction~\cite{Sternheimer:1950}.

Moving beyond elemental crystal structures, several compounds with seemingly bizarre stoichiometries (at least from the perspective of minds that experience a 1~atmosphere reality) have been predicted and/or synthesized under pressure. The familiar combination of sodium and chlorine, table salt and prototypical ionic compound with a 1:1 ratio, is not the only stable crystalline structure in the Na-Cl phase diagram. At least two unique stoichiometries, Na$_3$Cl and NaCl$_3$, were synthesized, and several others were predicted to become stable when squeezed~\cite{Zhang:2013}. Noble gases xenon, argon, and helium are active components of solid compounds that have been synthesized~\cite{Dewaele:2016b,Stavrou:2018,Dong:2017}, and \emph{very} hydrogen-rich compounds such as YH$_9$, LaH$_{10}$, and CaH$_6$ that are high-temperature superconductors with superconducting critical temperatures ($T_c$s) approaching room temperature have been made~\cite{Snider:2021,geballe2018synthesis,drozdov2019superconductivity,Somayazulu:2019-La,Ma:2022,Li:2022}. These metal superhydrides are materials-by-design success stories inspired by theoretical predictions ~\cite{Wang:2012,Liu:2017-La-Y,Peng:2017}.

In the following sections, we explore the plethora of exciting materials that can be created using high pressure, with all their intriguing structural and behavioral phenomena. While their stability and existence can be traced to the pressure driven electronic rearrangements of the constituent atoms described in Section~\ref{sec3}, the manifestations of these rearrangements can take many forms. Below, we describe a variety of illustrative phases sorted into a series of categories, but by necessity these categories will, at times, overlap. Nevertheless, all exemplify the ramifications of high pressure on solid-state chemistry.

\subsection{Electronic structure}\label{subsec5}

At atmospheric pressure, the stoichiometries of many inorganic solid-state compounds can be predicted from the most common oxidation states of their constituent elements. Usually,  alkali metals and alkaline earth metals possess oxidation states of +1 and +2 respectively, so when combined with the O$^{2-}$ ion one would expect Na$_2$O and MgO, as well as K$_2$O and SrO to form. The noble gases are mostly inert, the p-block is amenable to forming covalently bonded networks, and while a wide variety of oxidation states, which correspond to various filled or half-filled orbitals, are available to several transition metals, Zn, Cd, and Hg steadily persist in maintaining their $d^{10}$ configurations. 

The consequences of the orbital energy shifts discussed in Section~\ref{sec3} mean that several of these ``rules" no longer apply at high pressures, and elements can adopt unusual oxidation states in compounds with unexpected stoichiometries. The Li$_n$Cs phases~\cite{Botana:2014} used to illustrate the effects of pressure on electronegativity provide one such example. Above 70~GPa, the very lithium rich Li$_5$Cs phase is predicted to become stable, joined at higher pressures by Li$_3$Cs, Li$_4$Cs, and LiCs~\cite{Botana:2014}. Remarkably, the calculated Bader charges on the Cs atoms in these stoichiometries are all more negative than -1,  and since calculated Bader charges frequently underestimate formal oxidation states, in these compounds Cs may attain a formal oxidation state that is potentially lower than -2. While alkali metal anions (alkalides) have previously been captured at ambient pressures with cryptands~\cite{Dye:1979}, they achieve charges only up to -1 with the additional electron going into the $n$s orbital. In the case of the Li$_n$Cs phases, however, pressure induces significant Li 2s~$\rightarrow$~2p and Cs 6s~$\rightarrow$~5d electronic transitions, with the latter increasing progressively with higher Li content, thereby facilitating the acceptance of electron density by the typically unoccupied Cs 5d orbitals.  

\begin{figure}[!h]
\begin{center}
\includegraphics[width=\figurewidth]{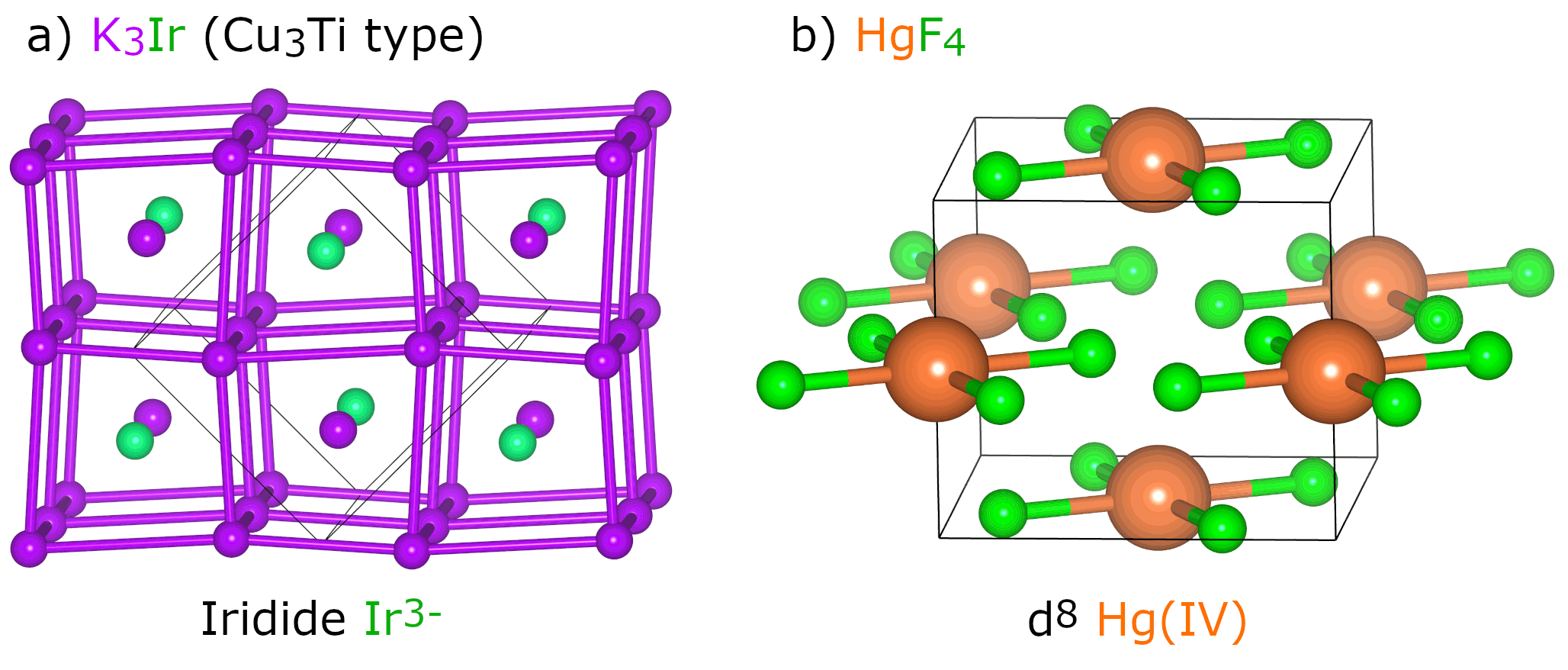}
\end{center}
\caption{High pressure compounds where the transition metal atoms adopt curious electronic configurations in (a) K$_3$Ir~\cite{Brgoch:2016} (Cu$_3$Ti type), with Ir@K$_8$ and K@K$_8$ distorted cubes and iridide I$^{3-}$ anions, and (b) HgF$_4$~\cite{Botana:2015}, in which d$^8$ configurations on the Hg lead to square planar geometries. 
\label{fig:K3Ir_HgF4}}
\end{figure}

For K, Cs, and Rb the pressure-driven $n\textrm{s} \rightarrow (n-1)\textrm{d}$ transitions led to predictions of transition-metal like behavior ~\cite{Bukowinski:1976,Pickard:2011} in the formation of intermetallic compounds with actual transition metals~\cite{Parker:1996,Hasegawa:1997,Lee:2003,Adeleke:2020}. Within some of these compounds, the transition metal elements assume exotic electronic configurations, as in the case of the predicted potassium iridide K$_3$Ir (Figure \ref{fig:K3Ir_HgF4}a) containing the Ir$^{3-}$ anion~\cite{Brgoch:2016}. The Ir 5$d$ orbital becomes fully occupied as a result of electron transfer from K, echoed in the later predicted Rb$_3$Ir and Cs$_3$Ir phases~\cite{Lotfi:2019}. K$_3$Ir and Rb$_3$Ir share the $Pmnm$ Cu$_3$Ti structure type, while Cs$_3$Ir adopts the $P2_1/m$ Ni$_3$Ta type, which consist of Ir@M$_8$ and M@M$_8$ distorted cubes, and Ir@M$_{12}$ distorted cuboctahedra respectively. In combination with Li under high pressure, Au displays a similar ability to adopt a significantly negative formal charge of less than -3 in the predicted phases Li$_4$Au and Li$_5$Au -- where electrons donated from Li are placed into the empty Au 6p orbitals~\cite{Yang:2016}, which are less destabilized than the Li 2s or 2p under pressure.

Pressure can also promote chemical interactions with core or semi-core orbitals, as in the case of HgF$_3$ and HgF$_4$~\cite{Botana:2015}. These stoichiometries are predicted to become stable above 73 and 38~GPa, respectively, and above 200~GPa HgF$_4$ is computed to decompose into HgF$_3$ and F$_2$. The $I4m$ symmetry HgF$_4$ crystal  possesses square planar HgF$_4$ units typical of a d$^8$ organometallic complex, with the Electron Localization Function (ELF)~\cite{Savin:2005} confirming covalent Hg-F interactions.  To form the four Hg-F bonds, not only the Hg 6$s$ but also two of the semicore 5$d$ electrons are required. In HgF$_3$, the $Fm\bar{3}m$ structure (which distorts below 100~GPa to $C2/m$ symmetry) involves a fluorite-type HgF$_2^+$ lattice stuffed with F$^-$ ions, leaving Hg with a d$^9$ configuration. A series of predicted CsF$_n$ phases, in which the Cs 5p electrons participate in Cs-F covalent bonds, are another example of the pressure-induced activation of core electrons~\cite{Miao:2013}. Their crystal structures display motifs resembling the isoelectronic [XeF$_n$]$^-$ molecules -- for example, $Fdd2$ CsF$_5$ contains planar pentagonal CsF$_5$ units, similar to the [XeF$_5$]$^-$ anion. With increasing fluorine content, the formal oxidation state on cesium reaches values greater than +1.

\subsection{High Pressure Electrides}\label{subsec6}

Electrides are solids where the electrons, localized on non nuclear-centric sites, behave as anions~\cite{Liu:2020b}. They are conceptually related to solvated electrons, in which the excess electrons can be thought to occupy cavities in the fluid~\cite{Zurek:2009}, as well as alkalide liquids \cite{Zurek:2011c,Zurek:2020h} or alkalide solids where alkali metal anions fill the interstitial voids~\cite{Dye:1979}. Although many types of electride families are known at atmospheric conditions, including those that are organic, inorganic, intermetallic and those where the electron localization is restricted to various dimensions or possesses topological properties, herein, we restrict the discussion to high pressure electrides: systems where the electron localization occurs as a response to compression~\cite{Miao:2015}. 

Though the formation of high pressure electrides has been rationalized in many ways, including pressure induced orbital rehybridizations~\cite{Neaton:2001,Ma:2009,Li:2010}, and multicenter bond formation~\cite{Dong_Oganov_Chapter,Modak:2019}, herein we focus on a simple model proposed by Miao and Hoffmann~\cite{Miao:2014,Miao:2015}.  As atoms in a solid compound are compressed, raising their orbital energies, the electrons in the highest-energy orbitals can vacate the atom entirely and occupy the interstices of the crystal lattice instead. To understand why this might occur, Miao and Hoffmann pointed out that orbitals can be ascribed to these voids, which can be thought of as interstitial quasiatoms (ISQs). At ambient pressure, the ISQ energies are higher than the atom-centered ones. However, unlike the atom-centered orbitals, those of the ISQ do not experience the repulsive effect caused by the core electrons, and their increase in energy with pressure will be less than the atom-centered orbitals. When the ISQ orbital energies fall below the atom-centered ones they will be occupied, thereby localizing the valence electrons in the interstitial regions. These electrons, detached from the nuclei, serve as anions and the corresponding compounds are called \emph{electrides}. 

\begin{figure}[!h]
\begin{center}
\includegraphics[width=\figurewidth]{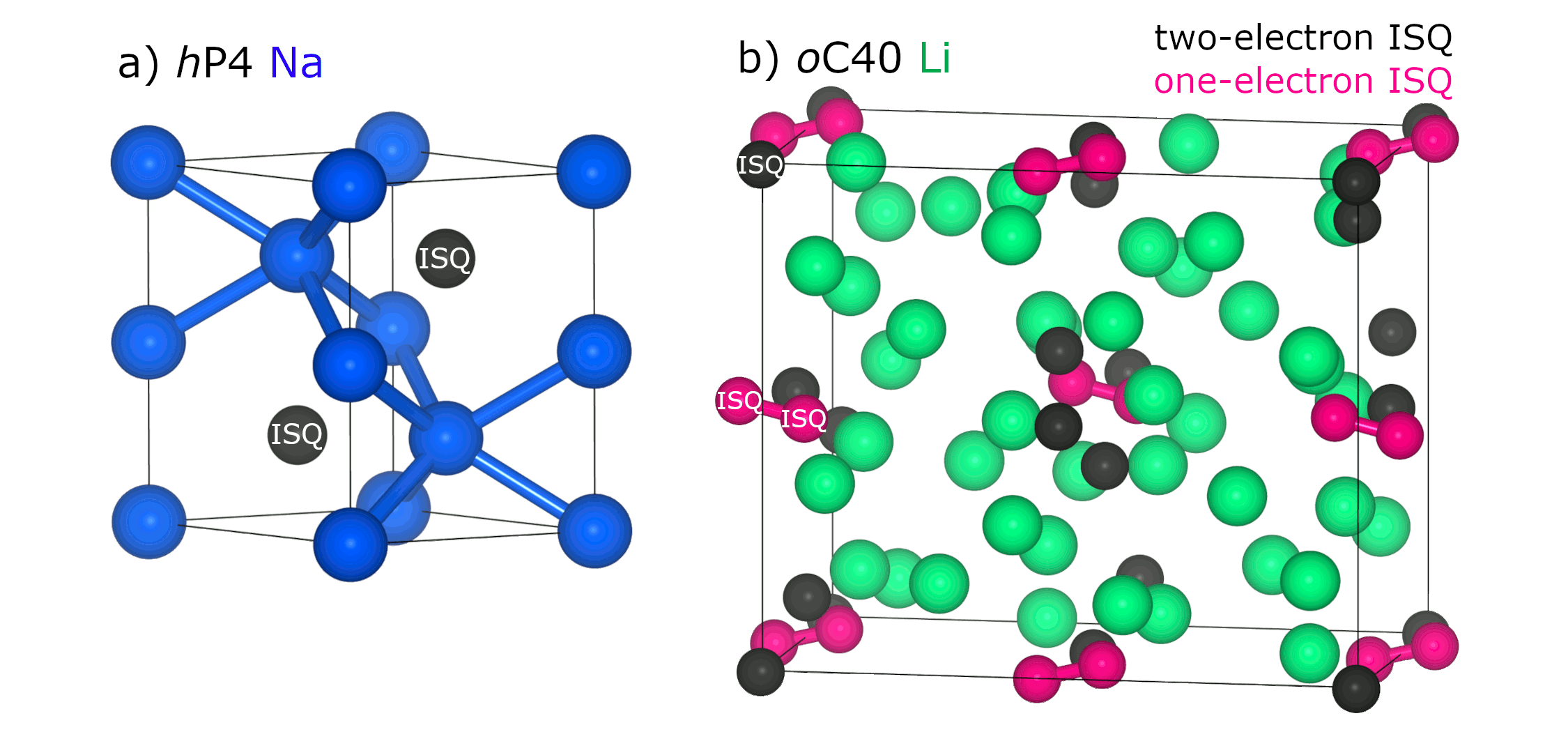}
\end{center}
\caption{High pressure electride formation is observed in a number of complex high-pressure phases of the alkali metals, including both the insulating (a) \emph{h}P4 phase of Na~\cite{Ma:2009} (Ni$_2$In structure type) where ISQs containing two electrons occupy the In sites, and (b) semiconducting \emph{oC}40 Li, containing three inequivalent ISQ sites of which two (black) are doubly occupied, and the third (pink) is singly occupied so a bond is formed between nearest neighbors, as highlighted by the lines that join them.
\label{fig:alkali_electrides}}
\end{figure}

For several simple metals, high-pressure phases identified as electrides via calculations have been subsequently studied experimentally. In sodium, Neaton and Ashcroft posited that under pressures high enough to induce overlap of the $2p$ orbitals, a combination of Pauli repulsion and core orthogonality constraints would drive the valence electrons away from the ionic cores, to localize in the crystalline interstices instead. This electronic redistribution would result in a metal-to-insulator transition~\cite{Neaton:2001}. Later CSP calculations -- with experimental confirmation of an optically transparent, wide-bandgap insulating phase in the same publication -- proposed an insulating $hP$4 phase with $P6_3/mmc$ symmetry to become stable above 260~GPa.~\cite{Ma:2009} The new $hP$4 phase was in fact experimentally observed at pressures as low as 200~GPa, but the discrepancy was ascribed to a combination of thermal effects as well as the preferential stabilization of metallic states by the computational method employed. Moving to higher pressures and temperatures, evidence for the $hP$4 phase has been obtained in shock-ramp experiments~\cite{Polsin:2022}. However, \emph{in situ} X-ray diffraction (XRD) revealed peaks that could not be attributed to  $hP$4 between 240-325~GPa at temperatures in the thousands-of-degrees Kelvin. Consistent with these dynamic compression experiments, calculations showed that the free energy of a $P6_3/m$ symmetry phase that is a topological electride was lower than that of $hP$4 at these pressures and temperatures~\cite{Wang:2022b}. At higher pressures yet, ca.\ 15.5~TPa, sodium is predicted to adopt a curious, metallic \emph{cI}24 electride phase consisting of Na$_{12}$ icosahedra~\cite{Li:2015}.  In the insulating $hP$4 structure that dominates much of the high-pressure landscape in sodium, highlighted in Figure~\ref{fig:alkali_electrides}a, the atoms occupy the Ni sites of the Ni$_2$In structure type with the ISQs on the In sites, in line with the treatment of this phase as (Na$^+$)$_2$E$^{2-}$ (where E$^{2-}$ denotes a doubly-occupied ISQ). In fact, several A$_2$X alkali metal chalcogenides adopt the antifluorite $Fm\bar{3}m$ crystal structure at ambient conditions, but eventually transition to the Ni$_2$In structure type under pressure~\cite{Vegas:2002,Santamaria-Perez:2011}. Potassium also adopts the $hP$4 phase when squeezed~\cite{Marques:2009,Degtyareva:2014}.

Lithium presents another example of complex structural and electronic behavior under pressure, as first postulated by Neaton and Ashcroft who suggested it may adopt an insulating, paired ground state~\cite{Neaton:1999}. Subsequent experiments revealed that Li assumes the same semimetallic $cI$16 \cite{Hanfland:2000} structure found in Na \cite{McMahon:2007}. At higher pressures, Li transitions to a number of unique phases, such as those with orthorhombic C-centered lattices and 88, 40, and 24 atoms in their unit cells which have been observed~\cite{Guillaume:2011}. One of these, $oC$40 with the $Aba2$ space group, is an electride displaying especially interesting behavior~\cite{Miao:2017}. In this phase, ISQs occupy three separate symmetry-distinct sites, two that are doubly occupied (E$^{\textrm{II}}$) and the third singly occupied (E$^\textrm{I}$), so its primitive cell can be considered as Li$_{20}$E$^{\textrm{II}}_8$E$^\textrm{I}_4$. The E$^\textrm{I}$-E$^\textrm{I}$ distance remains roughly constant at a short 1.3~\AA{} from 50-80~GPa, with an elevated electron density found between the ISQs. Crystal Orbital Hamilton Population (COHP)~\cite{Dronskowski:1993}, ELF, and  projected density of states (PDOS) analyses all indicate bonding character between the E$^\textrm{I}$ sites, and examination of the $\Gamma$-point band-decomposed charge densities revealed bonding and antibonding states analogous to the $\sigma_g$ and $\sigma_u^*$ orbitals in H$_2$. In fact, computations have shown that ISQs can form covalent, ionic and metallic bonds with atoms as well as with other ISQs~\cite{Miao:2015,Miao:2015b,Li:2010,Botana:2016,Miao:2017}. For this reason, Miao has espoused the idea that ISQs may be thought of as a chemical element and placed above helium in the periodic table under pressure \cite{Zurek:2019k}.



The proclivity of the elements to ISQ formation has been investigated by comparing the energies of their orbitals calculated at different pressures using a He confinement model with those of an ISQ 1s orbital~\cite{Miao:2014}. Unsurprisingly, Li and Na were found to favor ISQ formation at relatively low pressures, with Mg, Al, In, and Tl predicted to follow suit at higher pressures. Among the heavier alkali metals, the energies of the valence s orbitals were found to rapidly increase in energy relative to that of the ISQ 1s -- but as previously discussed, these elements are also susceptible to a pressure induced electronic $n\textrm{s} \rightarrow (n-1)\textrm{d}$ transition. Within cesium, for example, the increased d occupation, already noted by Sternheimer in 1950~\cite{Sternheimer:1950},  was invoked to explain the curious structure of Cs-IV adopted at 4.3~GPa, where the Cs atoms possess a coordination number of 8~\cite{Takemura:1982}. This decrease in coordination number with increasing pressure appeared so counterintuitive to Linus Pauling that he presented an alternative structure solution assuming cubic symmetry and invoking icosahedral clusters~\cite{Pauling:1989}. Pauling's hypothesis turned out to be incorrect. Importantly, von Schnering and Nesper~\cite{vonSchnering:1987} recognized that the Cs atoms in the Cs-IV structure occupied the Th sites of the ThSi$_2$ lattice -- and that the valence electron density exhibited maxima not near the Cs atoms but at the Si positions of ThSi$_2$ and between the Si-Si bonds, dubbing this phase an electride. This $I4_1/amd$ structure is also assumed by K and Rb under pressure~\cite{Olijnyk:1983,Ma:2008,Lundegaard:2009}. 

If the high-pressure phase behavior of the alkali metals were not complex enough, Na, K, and Rb all adopt different versions of an incommensurately modulated host-guest lattice (similar to the W$_5$Si$_3$ type), often referred to as the $tI$19 structure, a model of which is illustrated for K in Figure~\ref{fig:atompressure}d.~\cite{Olijnyk:1983,Schwarz:1999,McMahon:2001,McMahon:2006,McMahon:2006b,Gregoryanz:2008,Ma:2009,Woolman:2018} All three share the same host lattice, but display different periodicity in the guest lattice. In a study using commensurate approximants to model the electronic structure, electron localization in the interstitial spaces was found, with some highly localized basins as well as a 1D channel of electron density lying in channels of the host structure~\cite{Woolman:2018}. Another study proposed that electrides in the heavy alkali metals could be stabilized via ferromagnetic ordering~\cite{Pickard:2011}.

Numerous predicted binary and ternary systems under pressure also behave as electrides, including two Li$_3$Fe phases (with $P6/mmm$ and $P4/mbm$ symmetries)~\cite{Zhou:2016}, $P4/mbm$ Na$_3$Fe~\cite{Zhou:2016b}, and a superconducting Y$_3$Si phase~\cite{Zhang:2021}. In a range of Li$_n$I stoichiometry phases predicted above 50~GPa, ISQs form within the interstices between I-centered Li polyhedra -- but higher pressures drive electron density back from the ISQs to the iodine atoms, filling the 5p and eventually the 5d orbitals, skipping the 6s, in line with disfavoring s orbital occupation under pressure~\cite{Botana:2016}.  Finally, several electride phases have been identified involving the famously unreactive noble gases - including a particularly surprising case~\cite{Dong:2017} where the noble gas is crucial to the stability of the synthesized structure.

\subsection{Compounds of noble gases}\label{subsec7}

Where did all the xenon go? 

Relative to the abundance of Ar and Kr in the Earth's atmosphere, the amount of Xe is strikingly lower than it should be, a problem known as the ``missing xenon paradox". Geoscientists have explained this discrepancy in many different ways~\cite{Pepin:2002,Shcheka:2012,Marty:2012}, but a growing body of evidence suggests that the Xe did not escape, and instead it has been incorporated into the minerals found within the Earth. At Earth's core pressures, both the atomic orbital energies and the relative electronegativies of the elements, including Xe and the Fe and Ni that comprise the majority of the core, are significantly perturbed~\cite{Rahm:2019,Dong:2022}. Therefore, it should not be a surprise that their reactivity differs from our 1 atmosphere expectations.

Prior to the advent of widespread CSP, computational investigations concluded that Xe incorporation into Fe and Ni would not occur, at least not to a large extent~\cite{Caldwell:1997,Lee:2006}. These studies, which relied on the assumption that the Xe-metal alloys adopted crystal lattices similar to those of the elemental metals, turned out to be incorrect. Later CSP-based studies predicted the emergence of stable Xe-Fe and Xe-Ni compounds above 250 and 200~GPa, respectively~\cite{Zhu:2014}, with $Pm\bar{3}m$ XeFe$_3$ (AuCu$_3$-type) and $Pmmn$ XeNi$_3$ (based on Xe@Ni$_{12}$ cuboctahedra) having the lowest enthalpies of formation, although $P\bar{6}2m$ XeFe$_5$ and XeNi$_5$, as well as $P2_1/m$ XeNi$_6$ also appeared on the convex hull. Confirming these predictions, experimental studies have synthesized XeNi$_3$ and Xe(Fe/Ni)$_3$ phases at high pressure, although with slightly different structures than those predicted. This includes $Pm\bar{3}m$ for XeNi$_3$, either as an ordered AuCu$_3$~\cite{Dewaele:2016b} or disordered CrNi$_3$ alloy~\cite{Stavrou:2018}, and a mixture of $fcc$ and $Pmmn$-symmetry phases for XeFe$_3$~\cite{Stavrou:2018}. In these systems, the Fe and Ni atoms behave as oxidants, accepting 5p electron density from Xe~\cite{Zhu:2014,Stavrou:2018}, in agreement with the predicted increase in electronegativity differences between Xe and the transition metals at high pressure~\cite{Rahm:2019}. Xe$_2$FeO$_2$ and XeFe$_3$O$_6$, both involving substantial Fe-O and Xe-O bonding, have also been computed to be stable at pressures relevant to the Earth's core~\cite{Peng:2020}. The high-pressure ArNi phase, in which some Ni 3d electron density is transferred to the Ar 4s, inducing a magnetic moment on the Ni, has been synthesized~\cite{Adeleke:2019}.

CSP calculations have also predicted stable compounds containing Xe, or other noble gas (NG) elements, and Mg above 125-250~GPa ~\cite{Miao:2015b}. This includes Mg-Xe and Mg-Kr phases, which adopt structures based on stacked square lattices of Mg and the NGs in different patterns, ranging from the CsCl type ($Pm\bar{3}m$) to more complex $P4/nmm$ or $I4/mmm$ arrangements for MgNG and Mg$_2$NG stoichiometries. Compounds of Mg with Ar, on the other hand, were found to favor hexagonal arrangements such as anti-NiAs type MgAr ($P6_3/mmc$). In these compounds the energies of the metal 3s orbitals increase precipitously in comparison to the outer shell d orbitals of the noble gases inducing Mg 3s $\rightarrow$ NG d orbital transfer. The ELF of Mg$_2$NG (NG=Xe, Kr, and Ar) phases shows an additional interesting feature: ISQ formation.  This occurs far below the pressures at which elemental Mg is predicted to form an electride~\cite{Li:2010,Miao:2014}. Two reasons have been used to explain this phenomenon~\cite{Miao:2015b}. First, far fewer ISQ sites -- concomitantly occupying less space -- relative to the elemental Mg electride are necessary to accept the displaced valence electrons of Mg, as many of them are transferred to the NG atoms instead.  In addition, the NG atoms promote the formation of larger interstitial spaces in the structure, stabilizing the ISQ at lower pressures.
Under moderate pressures of ca.\ 50-300~GPa, the energies of the outer shell Xe d orbitals are similar to those of the ISQ 1s, although with higher pressure they become lower in energy, congruent with the gradual ISQ 1s $\rightarrow$ NG d electron transfer with increasing pressure up to 600~GPa~\cite{Miao:2015b}. 

Several other stable compounds of Xe have been predicted at high pressures, including XeO, XeO$_2$, and XeO$_3$~\cite{Zhu:2013}, as well as Xe$_3$O$_2$, Xe$_2$O, and Xe$_7$O$_2$~\cite{Hermann:2014}, while Xe$_2$O$_5$ and Xe$_3$O$_2$ have both been experimentally observed at pressures lower than 100~GPa.~\cite{Dewaele:2016} Krypton oxide, KrO, has been predicted as well~\cite{Zaleski-Ejgierd:2016}, as have xenon nitrides~\cite{Peng:2015,Howie:2016} and carbides~\cite{Bovornratanaraks:2019}. Fluorides of  argon~\cite{Kurzydlowski:2015}, krypton~\cite{Kurzydlowski:2017}, and xenon~\cite{Peng:2016} -- with Xe-Xe dimers cropping up in Xe$_2$F and XeF -- have all been predicted. Several xenon chlorides including XeCl, XeCl$_2$, and XeCl$_4$, with the former two being metastable by 10~GPa and reaching the convex hull by 60~GPa have been computationally studied~\cite{Zarifi:2018}. When they are combined with Li, the noble gases Ar~\cite{Li:2015b} and Xe~\cite{Liu:2017} are predicted to behave as anions, with the Li 2s orbital rising above the Xe and Kr outer shell d orbitals. Several cesium xenides are predicted to be stable at high pressures, many adopting alternate colorings of a distorted bcc lattice~\cite{Zhang:2015}. There is experimental evidence for the formation of a phase mixing Xe with water ice at conditions expected for planets such as Uranus and Neptune~\cite{Sanloup:2013}.

\begin{figure}[!h]
\begin{center}
\includegraphics[width=\figurewidth]{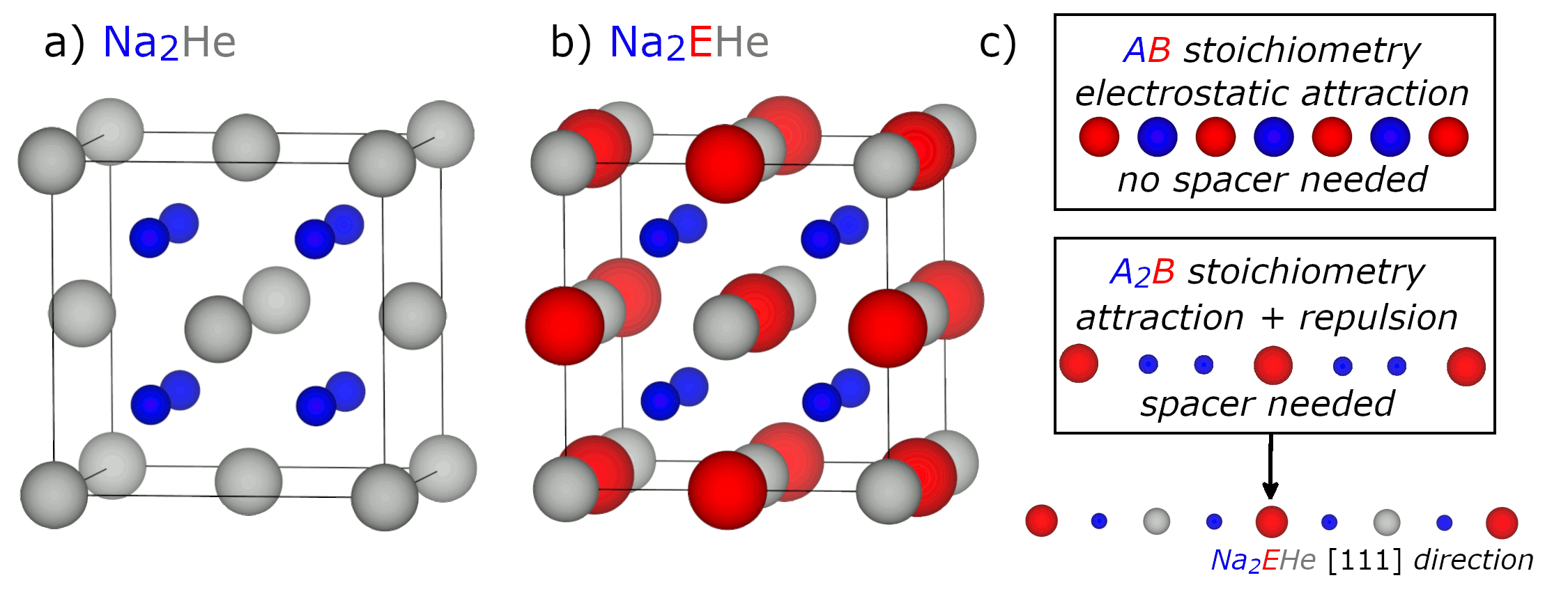}
\end{center}
\caption{Conventional unit cell of helium-containing Na$_2$He~\cite{Dong:2017} in the $Fm\bar{3}m$ space group (a), an electride with localized electron pairs occupying octahedral vacancies (b). The mechanism by which He incorporation serves to stabilize the structure is illustrated in (c), which shows schematically how He insertion reduces electrostatic repulsions involved in the A$_2$B (Na$_2$E) stoichiometry. 
\label{fig:Na2He}}
\end{figure}

Helium is famously the most inert element at ambient pressure by virtue of its closed-shell electronic configuration, zero electron affinity and large ionization potential. Nonetheless, a number of stable helium-containing compounds have recently been predicted at high pressure, including those with iron~\cite{Monserrat:2018}, ammonia~\cite{Liu:2020c}, water~\cite{Liu:2015b,Liu:2019}, nitrogen~\cite{Vos:1992,Li:2018d}, and even with \emph{other noble gases}~\cite{Wang:2015c,Cazorla:2009} (the van der Waals compound NeHe$_2$, a Laves phase in the MgZn$_2$ structure has been experimentally observed~\cite{Loubeyre:1993}). 

A particularly noteworthy example is provided by Na$_2$He~\cite{Dong:2017}, an electride phase with a fluorite-like lattice (Figure~\ref{fig:Na2He}a) in which every Na$_8$ cube that does not contain a He atom is instead occupied by an electron pair (Figure~\ref{fig:Na2He}b), so that the phase can be expressed as (Na$^+$)$_2$(E$^{2-}$)He. Although He does not participate in any bond formation, its presence is nevertheless a crucial stabilizing force in this phase, which has been successfully synthesized above 113~GPa~\cite{Dong:2017}.  A subsequent computational study showed that the He atoms act as inert ``spacers" to reduce the Madelung repulsion resulting from the unequal amounts of cations and anions in the parent (Na$^+$)$_2$(E$^{2-}$) phase~\cite{Liu:2018}. Reaction enthalpies for helium in combination with ionic AB, A$_2$B and AB$_2$ compounds revealed that He incorporation was generally favored when the cation:anion ratios were unequal such as in MgF$_2$ and Li$_2$O, but not for AB-type phases such as LiF or MgO, in line with the prediction of successful He incorporation into certain alkali oxides and sulfides~\cite{Gao:2019b}. Helium placement in the ionic lattices tends to separate ions of similar charge as shown schematically in Figure~\ref{fig:Na2He}c. An FeO$_2$He phase in which the Fe and O atoms form a fluorite lattice and the He atoms occupy the remaining Fe$_8$ cubes (isopointal to Na$_2$EHe) was predicted to be stable above 120~GPa~\cite{Zhang:2018}, with He appearing to play the same role of spacing agent. This mechanism allows even the most inert noble gases to play an active role in stabilizing compounds at high pressure, all without forming a single chemical bond.

\subsection{Miscibility under pressure}\label{subsec8}

The noble gases obtained their names due to their general lack of reactivity, but under ambient conditions, numerous combinations of elements resist mixing to form alloys or stoichiometric compounds. The proclivity or reluctance of a pair of elements towards mixing has been explained in a variety of ways, resulting in predictive rules including those of Hume-Rothery and co-workers~\cite{Hume-Rothery:1969}, Miedema's model~\cite{Miedema:1973,Miedema:1973b}, Darken and Gurry's maps~\cite{Darken:1953}, and more~\cite{Chelikowsky:1979,Alonso:1980,Wang:2015}.  As we will soon see, pressure turns out to be a useful variable that can alter the (im)miscibility of two or more elements. 

Consider, for example, magnesium and iron, whose large size and small electronegativity differential at ambient pressure makes compound formation intractable~\cite{Dubrovinskaia:2004}. According to Miedema's rules, compound formation is favored by large electronegativity differences and similar charge densities~\cite{Miedema:1973}. The electronegativity difference between the two elements greatly increases under pressure~\cite{Rahm:2019b,Dong:2022} -- and because Mg is more compressible, its radius approaches that of Fe when squeezed, thereby increasing the miscibility of the two elements. As a result, stable Mg-Fe compounds have been  computationally~\cite{Gao:2019,Fang:2021} and experimentally~\cite{Dubrovinskaia:2005} studied under pressure. The higher compressibility of K has also been found to favor compound formation with transition metals such as Ag, even at pressures below which K is anticipated to undergo an $n$s $\rightarrow$ ($n-1$)d transition~\cite{Atou:1996}.

\begin{figure}[!h]
\begin{center}
\includegraphics[width=\figurewidth]{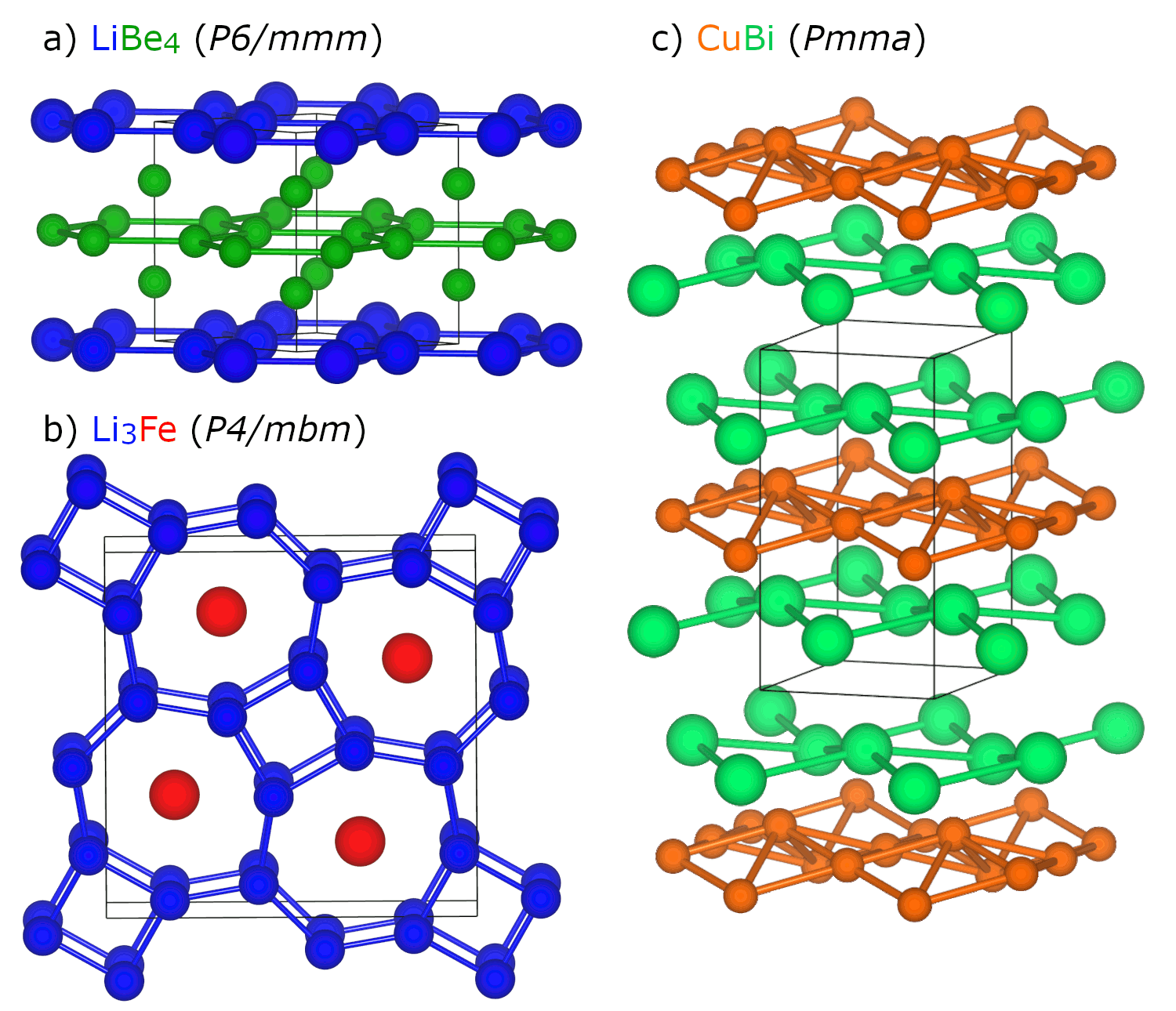}
\end{center}
\caption{High pressure enables the mixing of elements which do not form compounds or alloys otherwise. These include (a) LiBe$_4$, a layered compound of Li and Be~\cite{Feng:2008}, (b) Li$_3$Fe, a host-guest compound with Fe atoms as the guests to a Li-based host lattice~\cite{Zhou:2016}, and (c) CuBi~\cite{Guo:2017,Clarke:2017}, one representative of the many bismuth-containing compounds which have been found at high pressure. CuBi has been experimentally observed. 
\label{fig:miscibility}}
\end{figure}

Another case-in-point of pressure induced reactivity are the Li-Be alloys predicted to be stable in the megabar regime~\cite{Feng:2008}.  By 20~GPa LiBe$_2$ reaches the Li-Be convex hull, where it is joined by LiBe$_4$ (shown in Figure~\ref{fig:miscibility}a) and LiBe by 80~GPa. At 100~GPa the latter trades its place on the convex hull with Li$_3$Be. Alignment of strong diffraction peaks with 2$k_F$ (twice the free-electron Fermi wavevector) is suggestive of stabilization through a Fermi surface-Brillouin zone interaction mechanism, which has been used to explain the particular stabilities (and electron counts which make them so) of Hume-Rothery electron phases~\cite{Degtyareva:2006,Berger:2011}. Furthermore, at around 82~GPa, an odd feature emerges in the DOS curve of $P2_1/m$ LiBe: the base of the valence band appears as a step-like function, remains flat for $\sim$4~eV, and sharply increases once more before more complex features take over. This is linked to a distinct separation -- made possible by the pressure-induced increase in the electronegativity difference between Li and Be -- between high- and low-electron density planes associated with the Be and Li atoms, respectively, leading to quasi-2D-like behavior in a geometrically 3D structure. Stabilization through Fermi sphere interaction with higher zones (referred to as Jones zones from the Mott-Jones formulation of this mechanism) was invoked to propose a high-pressure NaAl phase in the NaTl structure type~\cite{Feng:2010} just above 12~GPa.

Another element that does not undergo compound formation with Li at ambient conditions is Fe. Just above 40~GPa, however, Li$_3$Fe ($P6/mmm$) and LiFe ($Fd\bar{3}m$, NaTl-type) phases are computed to lie on the convex hull~\cite{Zhou:2016}. Some interstitial electron localization appears in Li$_3$Fe, both in $P6/mmm$ symmetry as well as the $P4/mbm$ symmetry computed to prevail just above 60~GPa which is shown in Figure~\ref{fig:miscibility}b. Both Li$_3$Fe phases are host-guest lattices, with the Fe atoms lying in larger hexagonal ($P6/mmm$) and heptagonal ($P4/mbm$) channels, and the electron localization is found within the smaller triangular and square channels. Na$_3$Fe is predicted to adopt this same $P4/mbm$ phase between 120 and 300~GPa~\cite{Zhou:2016b}. An additional Li$_3$Fe$_2$ phase with $C2/m$ symmetry appears on the convex hull by 80~GPa, involving Fe zigzag chains in combination with alternating Li linear or armchair chains.

Recently, a host of bismuth-containing phases have been found to be stabilized at high pressure. Bismuth is a component in numerous topologically nontrivial materials~\cite{Chen:2009,Liu:2014} and superconducting systems at ambient pressure~\cite{Reynolds:1950,Matthias:1966,Hor:2010}, yet under these conditions does not form stable compounds with many of the transition metals. In combination with the exotic bulk properties of bismuth, the various magnetic or otherwise electronically nontrivial properties of the transition metals make for a tantalizing combination in high pressure experiments. Under pressure, the estimated electronegativity of Bi drops precipitously in comparison to most of the d-block, rendering it more reactive~\cite{Rahm:2019b,Dong:2022}. Indeed, above 32~GPa, FeBi$_2$~\cite{Walsh:2016,Amsler:2017} was observed to form in DAC experiments in the $I4/mcm$ Al$_2$Cu structure type shared by NiBi$_2$~\cite{Amsler:2018} and MnBi$_2$~\cite{Walsh:2019} (both are stabilized by high pressure although other Ni-Bi and Mn-Bi phases are accessible at atmospheric pressure), as well as certain high-pressure transition metal pnictides~\cite{Takizawa:1999,Wu:2009,Poffo:2012}. A second phase, FeBi$_3$ with $Cmcm$ symmetry, has also been predicted to lie on the convex hull between 36 and 39~GPa, but this narrow  stability range likely hinders synthetic accessibility~\cite{Amsler:2017}. 

In fact, high-pressure high-temperature methods have been used to synthesize a wide variety of bismuth-containing compounds including CoBi$_3$~\cite{Schwarz:2013,Tence:2014}, which adopts the $Pnma$ NiBi$_3$ structure type, becomes stable by 5~GPa, and is a superconductor with a $T_c$ just below 0.5~K. Synthesized binaries of Cu and Bi include Cu$_{11}$Bi$_7$ at 3~GPa~\cite{Clarke:2016} and CuBi (Figure~\ref{fig:miscibility}c) at 6~GPa~\cite{Guo:2017,Clarke:2017}, as well as a $I4/mmm$ Cu$_2$Bi phase above 50~GPa~\cite{Amsler:2017b}, which is possibly overtaken by a Cu$_7$Bi$_2$ phase~\cite{Amsler:2018}. Of the second-row transition metals, MoBi$_2$, also in the Al$_2$Cu structure type, has been synthesized above 35~GPa, while evidence for a Mo-Bi bcc-type alloy appeared above 5~GPa~\cite{Altman:2021}.

The Linear Approximation to Enthalpy (LAE), a tool for rapid and computationally cheap evaluation of formation enthalpies, was used to explore the high-pressure stability of structures in binary ambient-immiscible systems~\cite{Amsler:2018}. In concert with the minima hopping CSP method, several new phases were found to be stabilized by 50~GPa: PbAs, Si$_3$Al, SiAl, SiAl$_3$, BiSn$_3$ -- yet another bismuthide -- In$_3$Fe, Hg$_3$In, HgIn, HgIn$_3$, Hg$_3$Sn, ReSn$_3$, ReBr$_3$, ReGa, and ReGa$_3$. Only a limited range of stoichiometries (A$_3$B, AB, and AB$_3$) was sampled, encouraging further investigation into each of these systems -- but now there is preliminary data to suggest fertile ground.

\subsection{Geometries and bonding}\label{subsec9}

In the previous sections, we have explored curious electronic interactions made possible by external pressure and compound formation between unexpected species. Here, we shift our focus to the particular geometrical arrangements that emerge in materials under pressure. With higher density, atoms are forced into closer proximity -- the possibility of electride formation notwithstanding -- which can promote multicenter bonding in both electron-poor and electron-rich contexts as coordination numbers increase~\cite{Grochala:2007a}. When electron-precise species are closely bunched under compression, electron-deficient multicenter bonding can emerge as the constituent electrons are needed to span more bonding interactions.

\begin{figure}[!h]
\begin{center}
\includegraphics[width=\figurewidth]{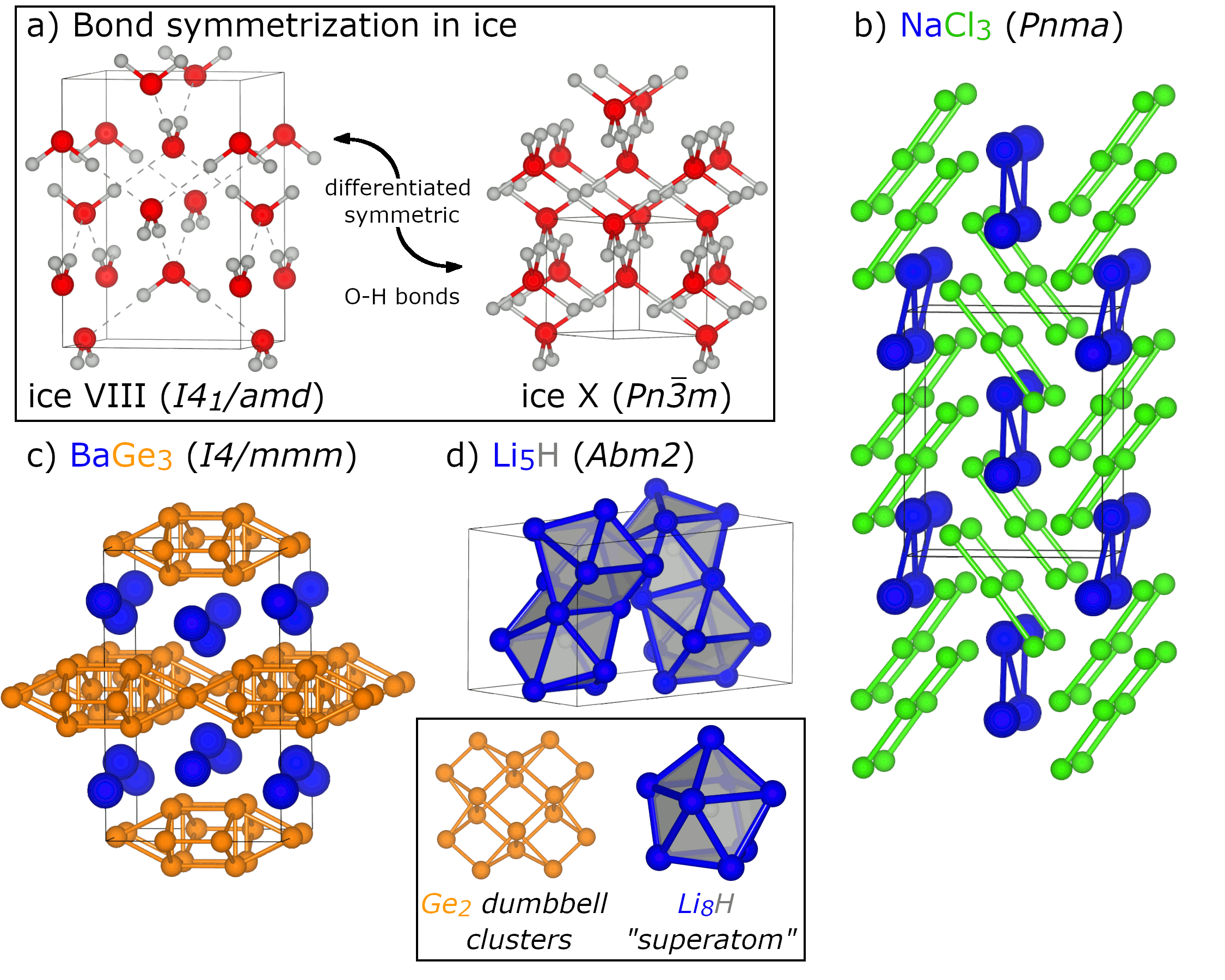}
\end{center}
\caption{Geometrical and bonding adaptations at high pressures. With pressure (a), covalent O-H and intermolecular hydrogen bonds between separate water molecules in the ice VIII phase equilibrate to yield the symmetric ice X phase~\cite{Goncharov:1996,Bernasconi:1998}. Pressure coincides with the appearance of strange motifs including (b) linear trichloride anions in $Pnma$ NaCl$_3$~\cite{Zhang:2013}, (c) clusters of Ge dumbbells (inset) in $I4/mmm$ BaGe$_3$~\cite{Zurek:2015,Castillo:2016}, and (d) Li$_8$H ``superatom-like" building blocks (inset) in $Abm2$ Li$_5$H~\cite{Hooper:2012c}. Except for Li$_5$H, all have been experimentally realized. 
\label{fig:geometries}}
\end{figure}

Bond symmetrization is a frequent secondary consequence of compression: an asymmetric fragment forced to occupy a progressively smaller space has less room for asymmetry. Often, this leads to a collapse into a symmetric and multicentered bonding regime, as was predicted for water ice under pressure by Pauling~\cite{Pauling:1960}. He suggested that the intermolecular hydrogen bonds between adjacent water molecules would shorten with pressure~\cite{Pauling:1960}, eventually becoming equivalent with the intramolecular O-H bonds, as illustrated in Figure~\ref{fig:geometries}a. This prediction was verified experimentally upon the discovery of ice X, where the lone pairs of the oxygen atoms are used to form additional covalent bonds, rendering the oxygen atoms tetrahedrally coordinated by hydrogens in a diamond-like network~\cite{Goncharov:1996,Bernasconi:1998}. Pressure-induced hydrogen bond symmetrization has also been observed in computations on hydrogen halide systems such as HF, HCl and HBr~\cite{Zhang:2010}, as well as in the record-breaking superconductor $Im\bar{3}m$ H$_3$S~\cite{Errea:2016}. 

Small homoatomic clusters alien to the 1 atmosphere pressure-trained mind are found or predicted for other elements in high pressure crystal structures. The wide structural variety has been explained by the increased stabilization of homonuclear bonds as compared to more polar or ionic bonds under pressure~\cite{Zurek:2019k}, favoring single-element clustering. An example of a compound containing novel homonuclear motifs is $Pnma$ NaCl$_3$ (Figure~\ref{fig:geometries}b), computed to be stable from 20 to 48~GPa, featuring a linear Cl$_3^-$ anion reminiscent of the more familiar  triiodide I$_3^-$~\cite{Zhang:2013}. Another such motif is the pentazolate N$_5^-$ ring, which can store more energy than the related azide anion  N$_3^-$, but is challenging to synthesize at ambient pressure~\cite{Vij:2002,Wang:2018}. This species, ubiquitous in high-pressure phases, is predicted to be a constituent of LiN$_5$~\cite{Shen:2015,Peng:2015a,Laniel:2018b,Zhou:2020} -- a phase that has been successfully quenched to ambient conditions after synthesis at 45~GPa~\cite{Laniel:2018} -- to sodium pentazolates NaN$_5$ and Na$_2$N$_5$~\cite{Steele:2015}, CsN$_5$~\cite{Peng:2015b}, CuN$_5$~\cite{Li:2018}, MgN$_{10}$ and BeN$_{10}$~\cite{Xia:2019}, ZnN$_{10}$~\cite{Liu:2020}, BaN$_5$ and BaN$_{10}$~\cite{Huang:2018}, SnN$_{20}$~\cite{Wang:2020}, and IrN$_7$~\cite{Du:2021}. Polynitrogen chains feature in many proposed high-pressure compounds of Cs~\cite{Peng:2015b}, Fe~\cite{Wu:2018}, Zn~\cite{Liu:2020}, Ba~\cite{Huang:2018}, Sn~\cite{Wang:2020}, Cd~\cite{Niu:2021}, Gd~\cite{Liu:2021}, and Ta~\cite{Bykov:2021}, the last of which has been experimentally realized. High pressure also facilitates the formation of silicon clusters in predicted phases including Si$_4$ squares in CaSi~\cite{Gao:2014}, extended networks and clathrate-like cages in silicides such as CsSi$_6$~\cite{Li:2018b}, MgSi$_5$~\cite{Hubner:2019}, and several lithium silicide compounds~\cite{Zhang:2016}.

Tetrel clusters comprise a family of polar intermetallic $I4/mmm$ symmetry compounds formed from alkaline earth or rare earth metals and group 14 tetrels in a 1:3 ratio. An example of these isotypic compounds, BaGe$_3$, is shown in Figure~\ref{fig:geometries}c. The clusters within it may be described as tetrel dumbbells condensed into cubes,  which are capped on four equatorial faces by additional dumbbells shared with a neighboring cube, forming a loose three-dimensional network. This structure has been experimentally observed in Ca, Y, and Lu silicides~\cite{Schwarz:2012} (and later identified in a CSP investigation of the Y-Si system~\cite{Zhang:2021}), and a related distorted $I\bar{4}2m$ BaSi$_3$ phase has also been synthesized~\cite{Hubner:2019b}. Alkaline earth trigermanides CaGe$_3$~\cite{Schnelle:2012}, SrGe$_3$~\cite{Nishikawa:2015}, and BaGe$_3$~\cite{Zurek:2015,Castillo:2016} have also been found to adopt this structure.  The distribution of electrons within the clusters aligns with two-center two-electron bonds along the tetrel dumbbells, with multicenter interactions between the tetrel and rare earth/alkaline earth~\cite{Schwarz:2012,Zurek:2015,Castillo:2016b}. Superconductivity has been measured in some of these compounds, albeit at low temperatures~\cite{Schwarz:2012,Castillo:2016}, augmented by predictions from first principles calculations~\cite{Zurek:2019k,Castillo:2016b,Zhang:2021}. Additionally, the stability conferred by the strong covalently bonded networks permits some of these phases, synthesized at high-temperatures and high-pressures, to be recovered at ambient conditions~\cite{Castillo:2016b,Hubner:2019b}. A similar example is presented by elemental carbon -- diamond is its ground state at high-pressures, but due to the immense strength of its sp$^3$ covalently bonded network, the energetic barrier for its transition to the lower-enthalpy allotrope graphite is too high and it persists ``forever'' under ambient conditions. Furthermore, laser-driven ramp compression studies of carbon to 2~TPa have found that carbon stubbornly maintains the diamond structure well beyond its predicted high-pressure stability limits~\cite{Lazicki:2021}, as the barriers to breaking the sp$^3$ bonds remain large under pressure.

Another example of unique clusters predicted to form only at high pressures are found within a family of lithium subhydrides~\cite{Hooper:2012c}.  Computations uncovered two nearly isoenthalpic  Li$_5$H phases that had the most negative enthalpies of formation. Both were built of Li$_8$H units that behaved as superatoms analogous to similar units in synthesized Rb$_9$O$_2$ and Cs$_{11}$O$_3$ suboxides~\cite{Simon:2010}. One of these, with $Abm2$ symmetry, is shown in Figure~\ref{fig:geometries}d. The Li$_8$H cluster, a distorted bicapped trigonal antiprism of Li encapsulating a single H atom, has one electron in excess of the closed-shell octet and thus behaves as a superalkali atom. 

In addition to the well-known H$_2$ molecular units and H$^-$ hydridic species, hydrogen atoms can form other distinct clusters. One of these, the trihydrogen cation, H$_3^+$, is in fact one of the most abundant species in the universe -- but it is also largely relegated to interstellar space and the atmospheres of gas giant planets~\cite{Oka:2013}. High pressure crystal chemistry offers another opportunity. The halogen polyhydride $Cc$ H$_5$Cl~\cite{Wang:2015b,Duan:2015,Zeng:2017}, predicted to become stable above 100~GPa, contains slightly distorted H$_3^+$ clusters with H-H distances of 0.74, 0.97 and 1.01~\AA{}~\cite{Wang:2015b}. By 300~GPa the three distances converge to 0.87-0.88~\AA{}, yielding a nearly-perfect equilateral triangle. With sufficient pressure, this H$_3^+$ unit interacts with a neighboring H$_2$ molecule forming a twisted bowtie-like loosely interacting H$_5^+$ motif~\cite{Wang:2015b,Zeng:2017}. Metastable predicted H$_2$F, H$_3$F, and H$_5$F species, as well as H$_5$Br also contain this triangular H$_3^+$ cation~\cite{Duan:2015}, as does the metastable $P1$ LiF$_4$H$_4$~\cite{Bi:2021}. 

With two extra electrons, the trihydride anion, H$_3^-$, prefers a linear arrangement involving a three-center four-electron bond. Quantum chemical calculations have shown that the ground state geometry of the isolated trihydride anion possesses one H-H bond that is substantially longer than the other (2.84 vs. 0.75~\AA{}), with the transition state between the H-H$\cdot\cdot\cdot$H and H$\cdot\cdot\cdot$H-H configurations corresponding to the symmetric case~\cite{Ayouz:2010}. Nevertheless, certain predicted high-pressure hydrides of the heavy alkali metals K~\cite{Hooper:2012}, Rb~\cite{Hooper:2012a}, and Cs~\cite{Shamp:2012} (as well as the alkaline earth metal Ba~\cite{Hooper:2013}) feature an H$_3^-$ anion symmetrized via pressure. Synthesized NaH$_7$ is thought to contain an asymmetric linear H$_3^-$ motif~\cite{Struzhkin:2016}. Linear H$_3^-$ units are also predicted to appear in various indium~\cite{Liu:2015} and lithium polyhydrides~\cite{Chen:2017}.

Scandium polyhydrides, meanwhile, are predicted to feature five-membered rings of hydrogen atoms in various arrangements. In $I4_1/md$ ScH$_9$, which lies on the convex hull around 300~GPa~\cite{Zurek:2018b}, strips of edge-sharing H$_5$ pentagons are stacked perpendicular to one another along the \emph{c} axis, linked by vertices. The strips are separated by additional H atoms in molecular H$_2$ units. Around 250~GPa, ScH$_{10}$ adopts a $Cmcm$ structure in which H$_5$ pentagons are grouped into sets of three, sharing edges and a single common vertex. This phase is nearly isoenthalpic with another ScH$_{10}$ structure with the same (H$_5$)$_3$ ``pentagraphenelike" clusters but arranged in $P6_3/mmc$ symmetry~\cite{Xie:2020}. ELF demonstrates bonding within the H$_5$ units~\cite{Zurek:2018b,Xie:2020}. This same pentagraphenelike structure is expected to lie on the Lu-H convex hull at 300~GPa and to be very near the Hf-H and Zr-H hulls~\cite{Xie:2020}. For higher hydrogen contents yet, ScH$_{12}$ is predicted to be built of stacked strips of edge-sharing H$_5$ pentagons spaced by Sc~\cite{Zurek:2018b}. 

\begin{figure}[!h]
\begin{center}
\includegraphics[width=\figurewidth]{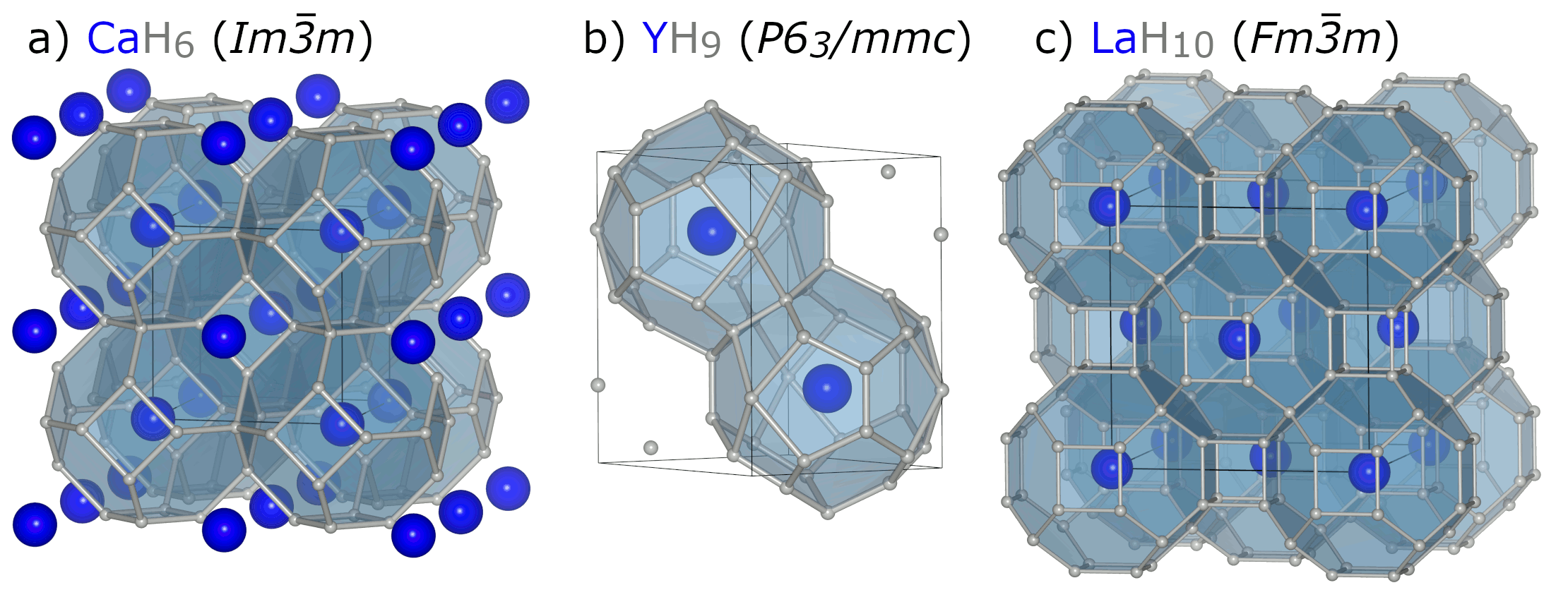}
\end{center}
\caption{Clathrate-like metal hydrides predicted then synthesized at high pressures. Their remarkable superconducting properties are tied to strong electron-phonon coupling. Several metal-hydrogen stoichiometries adopt these so-called ``superhydride'' motifs, including $Im\bar{3}m$ MH$_6$, exemplified by CaH$_6$~\cite{Wang:2012,Li:2022,Ma:2022}, (b) $P6_3/mmc$ MH$_9$, exemplified by YH$_9$~\cite{Peng:2017,Snider:2021}, and (c) $Fm\bar{3}m$ MH$_{10}$, exemplified by LaH$_{10}$~\cite{Liu:2017-La-Y,Peng:2017,drozdov2019superconductivity,Somayazulu:2019-La}. 
\label{fig:clathrate_hydrides}}
\end{figure}

Indeed, the first, most simple, element does not like to be outdone! One more class of high-pressure hydrogen rich materials whose prediction sparked tremendous experimental synthetic efforts are the so-called ``metal superhydrides''. The reason why scientists have pursued them in earnest is the prediction, verified by recent experiments, of conventional superconductivity at temperatures approaching those experienced in a cold room (the $T_c$ of LaH$_{10}$ is about 10~$^\circ$F -- January in Siberia), or a crisp fall day (the $T_c$ of C-S-H is about 60~$^\circ$F), albeit still at very high pressures! CSP investigations into the high-pressure Ca-H system revealed a curious $Im\bar{3}m$ CaH$_6$ phase (Figure~\ref{fig:clathrate_hydrides}a) stable above 150~GPa in which the Ca atoms were arranged in a bcc lattice and the H atoms condensed into a sodalite-like H$_{24}$ framework~\cite{Wang:2012}. All H-H distances were equivalent, 1.24~\AA{} at 150~GPa, and ELF analysis confirmed weak covalent bonding between the H atoms. This phase, a good metal, was predicted to exhibit large electron-phonon coupling, and indeed first principles calculations estimated a superconducting transition temperature $T_c$ of 220-235~K at 150~GPa. Subsequent synthetic exploration led to measurement of $T_c$ over 200~K at 160-170~GPa for CaH$_6$~\cite{Ma:2022,Li:2022}.

The computational discovery of CaH$_6$ was shortly followed up by fruitful theoretical investigations into related metal-hydrogen systems, turning up isostructural phases for Mg~\cite{Feng:2015a}, Sc~\cite{Qian:Sc-2017,Abe:Sc-2017,Zurek:2018b}, Y~\cite{Li:2015a,Liu:2017-La-Y,Peng:2017}, Pu~\cite{Zhao:2020}, Tb~\cite{Hai:2021}, Eu~\cite{Semenok:2021}, and Pm-Lu~\cite{Sun:2020b}. Structures that are distortions of this high symmetry phase have been predicted as well. This includes a tetragonally-distorted $I4/mmm$ ZrH$_6$ variant~\cite{Abe:2018-Zr}, along with an $R\bar{3}m$ phase in which opposite hexagonal faces of the H$_{24}$ cubic sodalite framework are opened  for SrH$_6$~\cite{Hooper:2013b,Wang:2015a,Zurek:2018d} and LaH$_6$~\cite{Liu:2017-La-Y,Peng:2017}, although the latter may not lie on the convex hull. $Imm2$ BaH$_6$~\cite{Hooper:2013}, which contains some of the H$_3^-$ trihydride anions explored above, can be thought of as a highly fragmented version of the sodalite framework. The role of distortions of the high-symmetry $Im\bar{3}m$ structure adopted by CaH$_6$ tends to reduce the density of states at the Fermi level, $E_F$, thereby also lowering $T_c$. The origin of such distortions has recently been investigated using the lens of DFT-Chemical Pressure~\cite{Hilleke:2022}.

Higher hydrogen content allows for other clathrate-like arrangements of hydrogen, from $P6_3/mmc$ YH$_9$~\cite{Peng:2017} (Figure~\ref{fig:clathrate_hydrides}b) to $Fm\bar{3}m$ LaH$_{10}$~\cite{Liu:2017-La-Y,Peng:2017} (Figure~\ref{fig:clathrate_hydrides}c). Distorted versions of these two structures have also been predicted, including $C2/m$ CaH$_9$~\cite{Shao:2019b} and $P1$ Eu$_4$H$_{36}$~\cite{Semenok:2021}, as well as $R\bar{3}m$ CaH$_{10}$~\cite{Shao:2019b} and AcH$_{10}$~\cite{Semenok-2018}. In the case of LaH$_{10}$, quantum anharmonic effects were found to be key in stabilizing the $Fm\bar{3}m$ structure over less symmetric variants~\cite{Errea:2020}. Other more complex clathrate-like hydrogenic frameworks have been predicted as well. One example is a diamond-like lattice of Mg@H$_{28}$ clusters intercalated with Li@H$_{18}$ units, which comprise the $Fd\bar{3}m$ Li$_2$MgH$_{16}$ phase. This compound is an example of a ``hot'' superconductor whose estimated $T_c$, 351~K at 300~GPa, is well above room temperature~\cite{Sun:2019}. Such materials are under intense speculation and investigation for their promise towards achieving room-temperature superconductivity, as described in Section \ref{subsec12} below.

\section{Superconductivity}\label{sec5}

\begin{figure}[!h]
\begin{center}
\includegraphics{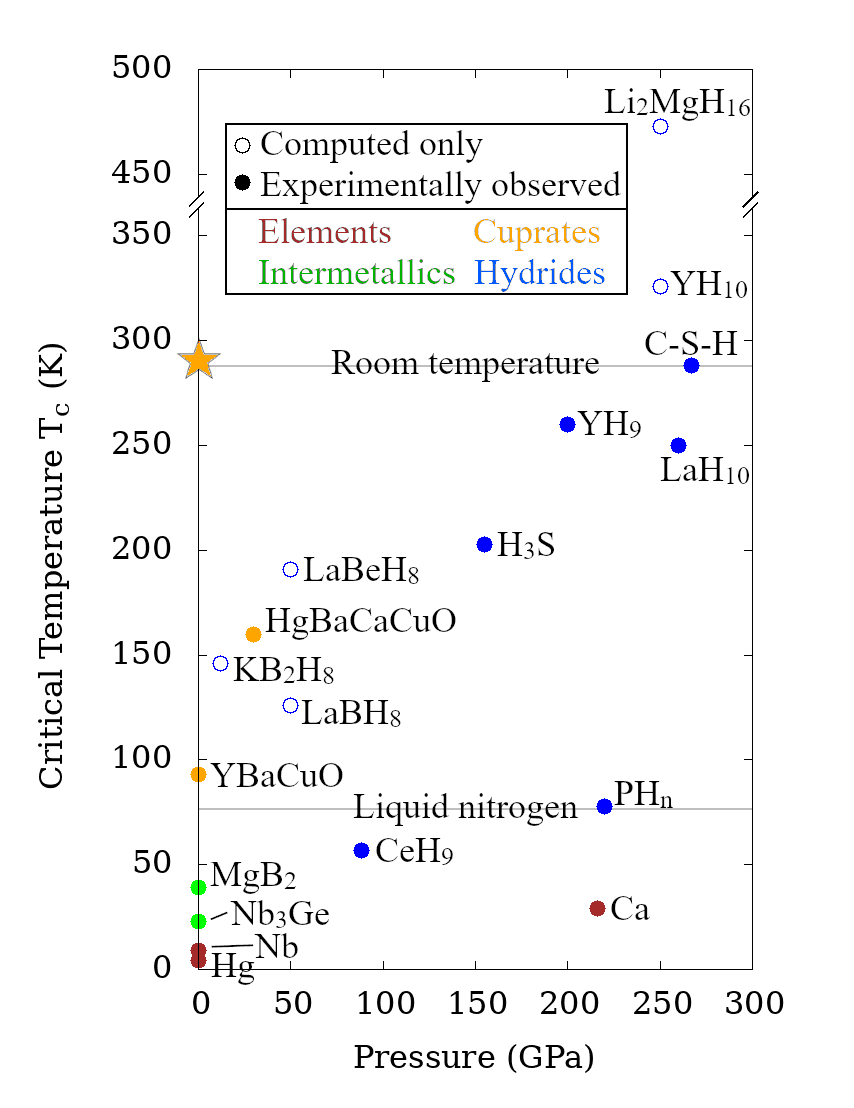}
\end{center}
\caption{Critical superconducting temperature, $T_c$, plotted against pressure for selected materials. Data points based on experimental measurements of $T_c$ are plotted with filled circles, while those estimated via theoretical calculations are plotted with empty circles. Elemental $T_c$s are given in brown, intermetallics in green, cuprates (belonging to the non-BCS unconventional family) in orange, and hydrides in blue. The boiling point of liquid nitrogen and room temperature are provided to guide the eye, with a star marking the ``holy grail" of room-temperature superconductivity at ambient pressure. Data was collected from references~\cite{Onnes:1911,Finnemore:1966,Gavaler:1973,Wu:1987,Gao:1994,Nagamatsu:2001,Sakata:2011,Drozdov:2015,Drozdov:2015a,Liu:2017-La-Y,Peng:2017,Somayazulu:2019-La,drozdov2019superconductivity,Salke:2019,Sun:2019,Snider:2020,diCataldo:2021,Gao:2021,Zhang:2022}.
\label{fig:Tc_vs_P}}
\end{figure}

The 1911 discovery of a phenomenon in which a substance's resistivity can plummet to zero~\cite{Onnes:1911} sparked countless investigations and resulted in a Nobel prize for Heike Kamerlingh Onnes, as well as a number of future Nobel prizes (directly or indirectly). The mechanism of superconductivity, and the search for new superconducting materials, has fascinated scientists for over a century. A key parameter for superconductors is the critical temperature, $T_c$, below which a material becomes superconducting. For a number of illustrative superconducting materials, $T_c$ is plotted against the pressure at which they are stable in Figure~\ref{fig:Tc_vs_P}. Of course, practical applications of superconductivity are limited if temperatures very near 0~K are required, and for some decades the highest known $T_c$ values lingered in the low twenties~\cite{Matthias:1954,Gavaler:1973}. This sparked debate regarding a possible natural ``cap" on superconductivity around these temperatures~\cite{AndersonMatthias:1964}. However, a family of cuprates whose superconducting mechanism has yet to be explained were the first to break the liquid nitrogen barrier~\cite{Wu:1987}, achieving $T_c$s over 160~K in the case of pressurized HgBa$_2$Ca$_{m-1}$Cu$_m$O$_{2m+2+\delta}$~\cite{Gao:1994}.

The 2001 discovery of superconductivity up to 39~K in MgB$_2$~\cite{Nagamatsu:2001} provided experimental evidence that non-cuprate phases could be promising superconductors, despite the fact that they deviated from the collection of empirical rules enumerated by B.\ T.\ Matthias in the 1950s and 1960s~\cite{Matthias:1957}. These ranged from favorable valence electron concentrations to a general distrust for theorists. MgB$_2$ belongs, with early and long-term record holders Nb$_3$Sn and Nb$_3$Ge, as well as the clathrate-like hydrides discussed in Section~\ref{subsec9}, to the family of ``conventional" superconductors whose mechanism is thought to follow the theory propounded by Bardeen, Cooper, and Schrieffer in 1957~\cite{bcs1,bcs2}. This discovery revitalized interest in conventional superconductors~\cite{Pickett:2007}, leading researchers to wonder what the trajectory of superconductivity research would have looked like had the 1957 measurements of the heat capacity of MgB$_2$~\cite{Swift:1957} captured the discontinuity that appears upon the superconducting transition.

Within BCS theory, two electrons of opposite momentum and spin that are within $\pm\hbar\omega_\text{cut}$ (an energy in line with the phonon energies) of the Fermi surface may, at long distances, overcome Coulombic repulsion and experience a net attractive potential when the lattice is polarized through phonon vibrations. This forms a Cooper pair, a composite boson of weakly interacting species. Thermal energy can break the Cooper pairs, with $T_c$ describing the temperature at which this occurs and the superconducting state is destroyed. From this construction, the $T_c$ for a material can be estimated by
\begin{equation}
 k_BT_{c} = 1.14\hbar\omega\exp\bigg[{\frac{-1}{N_FV}}\bigg] 
\label{eq:tc1} \end{equation}
where $\omega$ is the average phonon energy, $N_F$ is the single spin electronic density of states at $E_F$, and $V$ is the pairing potential between two electrons resulting from the electron-phonon interaction. This suggests that high $T_c$ is correlated with a large $N_F$ (a feature potentially tunable via judicious doping),  strong coupling between electrons and phonons, and high phonon frequencies. A frequently used semiempirical formula to estimate the $T_c$ of a conventional superconductor is the Allen-Dynes modified McMillan equation~\cite{McMillan:1968,Dynes:1972,Allen:1975}:
\begin{equation}
  T{_c} = \frac{\omega_\text{ln}}{1.2} \text{exp}\bigg[-\frac{1.04(1+\lambda)}{\lambda-\mu^*(1+0.62\lambda)}\bigg] .
\label{eq:ADM} \end{equation} 
where $\lambda$ is the electron-phonon coupling constant, $\omega_\text{ln}$ is the logarithmic average phonon frequency and $\mu^*$ is the Coulomb repulsion parameter, which is typically treated semiempirically. An approximate -- and illustrative -- formula to estimate $\lambda$ was proposed by Hopfield as~\cite{Hopfield:1969}
\begin{equation}
 \lambda = \frac{N_F\langle I^2 \rangle}{M \langle \omega^2 \rangle}
\label{eq:lambdahopfield} \end{equation}
where $\langle I^2 \rangle$ are the electron-phonon matrix elements averaged over the Fermi surface, $M$ is the atomic mass, and $\langle \omega^2 \rangle$ the mean phonon frequency. To increase $\lambda$, then, $N_F$ and the electron-phonon matrix elements should be increased. Converse to expectations from Equation~\ref{eq:tc1} and Equation~\ref{eq:ADM}, where the average phonon energy was directly proportional to $T_c$, here an increase in $\langle \omega^2 \rangle$ serves to decrease $\lambda$. In the denominator of Equation~\ref{eq:lambdahopfield}, $\langle \omega^2 \rangle$ and $M$ will naturally tend to counteract one another, as increases in atomic mass lead to softer phonon frequencies. In fact, evidence suggests that $T_c$ frequently increases at the edge of dynamic instability, where soft phonons promote strong coupling~\cite{Quan:2019}.

In the following sections, we describe the effect of pressure on the propensity for superconductivity of the elements, hydrogen in particular. We end by discussing families of hydrogen-rich phases that are extremely promising towards achieving the once-distant goal of room-temperature superconductivity (albeit potentially only at very high pressures).

\subsection{The elements}\label{subsec10}

Most of the elements in the periodic table can be superconducting given the right conditions. This includes, so far without exception, rather low temperatures. However, external pressure can affect the $T_c$, or even induce superconductivity in some elements~\cite{Buzea:2004,Schilling:2006,Hamlin:2015,Shimizu:2015}. In fact, calcium at ambient pressure is not a superconductor and achieves $T_c=$~29~K at 216~GPa~\cite{Sakata:2011}. In comparison, the highest elemental $T_c$ at ambient pressure is 9.2~K for niobium~\cite{Finnemore:1966}, while only lithium among the alkali metals~\cite{Tuoriniemi:2007}, and only beryllium~\cite{Falge:1967} among the alkaline earth metals are known to superconduct at ambient pressure, both well under 0.1~K. Of the fifty-four known superconducting elements of the periodic table, only thirty-one are superconductors at ambient pressure. 

Early indications regarding the ability of pressure to either enhance or suppress superconductivity~\cite{Onnes:1925,Jennings:1958,Smith:1967} were less than encouraging. For simple metals whose electronic structure aligns with a mostly free-electron model, such as Zn, Cd, Hg, and the group 13 metals, applied pressure serves to suppress what superconductivity is present at ambient pressure~\cite{Jennings:1958,Smith:1967}. In such free-electron-like metals, the effect of pressure is to broaden electronic bands and increase phonon frequencies due to the stiffer lattice. Band broadening reduces the electronic density of states at $E_F$, while a stiffer lattice is correlated with weaker coupling between electrons and phonons, both effects being detrimental to the superconductivity of a system. At ambient pressure, the alkali metals behave as free-electron metals, but as we have seen in Section~\ref{subsec6}, with pressure their electronic structure rapidly diverges from these expectations.

For Li, $T_c$ is highly pressure-dependent and reliant on complex crystal chemistry~\cite{Struzhkin:2002,Shimizu:2002,Deemyad:2003}. Like all of the alkali metals, it adopts a bcc crystal structure at ambient pressure, which in short order transitions to the fcc structure. Up to 8~GPa, the increase in pressure is reflected in an increase in phonon frequencies, typical for a stiffening lattice, but with further pressure the phonons become softer and just above 30-40~GPa, imaginary modes related to a structural instability appear. This pressure corresponds to another phase transition, this time to the \emph{h}R1 ($R\bar{3}m$) structure, and shortly thereafter to the \emph{c}I16. Maxima in $T_c$ are associated with the onset of dynamical instability, as the very soft phonon motions strongly bolster the electron-phonon coupling~\cite{Quan:2019,Chen:2020}. From its mostly spherical character at 0~GPa, where lithium adopts a bcc crystal structure, the pressure-induced 2s~$\rightarrow$~2p electronic transition drives an increasingly anisotropic Fermi surface featuring \emph{hot spots} of especially strong coupling~\cite{Kasinathan:2006}, losing free-electron-like behavior and leading to Fermi surface nesting (FSN)~\cite{Profeta:2006,Rodriguez-Prieto:2006,Xie:2007}. Phonon softening, in particular along the $\Gamma \rightarrow \textrm{K}$ path, is accompanied by enhancement of the electron-phonon coupling ~\cite{Yao:2009b,Bazhirov:2010}, with the result that $T_c$ grows from practically zero to a maximum of $\sim$20~K. This value is among the higher elemental $T_c$s, as a result of the pressure-induced electronic transitions in lithium. Following the structural transition to the \emph{c}I16 phase, lithium remains superconducting but $T_c$ decreases due to a reduction of the FSN and concomitant smaller electron-phonon coupling~\cite{Yao:2009b,Rousseau:2011,Yue:2018}. At high enough pressures, lithium undergoes a metal-semiconductor transition -- and eventually goes back to being a (poor) metal upon transitioning to the \emph{o}C24 ($Cmca$) phase~\cite{Matsuoka:2014}, but one nonetheless predicted to be superconducting with an estimated $T_c$ of 14~K at 200~GPa~\cite{Yan:2016}. 

In cesium~\cite{Wittig:1970,Schwarz:1998} and rubidium~\cite{Deng:2019,Schwarz:1999}, the onset of superconductivity with pressure is associated with the \emph{o}C16 (Cs-V and Rb-IV) structures, alongside the $\textrm{n}s \rightarrow (\textrm{n}-1)d$ transition~\cite{Deng:2019}. An increase in \emph{d}-character in the electronic states at $E_F$  is correlated with higher $T_c$ in the transition metals~\cite{Debessai:2008,Slocombe:2015} and applies here as well to the heavier alkali metals -- which, as we have seen, behave under pressure as transition metals themselves.

\subsection{Hydrogen}\label{subsec11}

Vitaly Ginzburg, awarded the Nobel Prize in Physics in 2003 for his work in superconductivity and superfluidity, formulated a list of, in his view, the thirty most pressing problems for physics in the 21st century~\cite{Ginzburg:1999,Ginzburg:2003}. Following controlled nuclear fusion, the second and third items on this list were high-temperature (room-temperature) superconductivity and metallic hydrogen. These problems are not unrelated.

In 1926, J.\ D.\ Bernal proposed that under sufficient pressure hydrogen would transition to a metallic state, but it took nearly a decade for pen to be put to paper by  Wigner and Huntington~\cite{Wigner:1935}. Their 1935 suggestion that hydrogen could be metallized by 25~GPa, estimated using a series of assumptions regarding crystal structure and compressibility, proved to be an immense underestimate. In 1968 Neil Ashcroft explicitly linked the quest for metallic hydrogen with the quest for superconductors with higher $T_c$s, with the suggestion that metallic hydrogen itself would be quite a fantastic superconductor~\cite{Ashcroft:1968}. Hydrogen, being the lightest element, can possess the highest frequencies (as a diatomic molecule) and experience a large electron-phonon coupling due to the lack of screening by core electrons. Moreover, in the metallic state its DOS at $E_F$ is thought to be quite high,  making for a very attractive material.

The pressure required to metallize hydrogen, however, is in the multimegabar range. Claims of metallic or semimetallic hydrogen have been made for DAC experiments at very high pressures and low temperatures, toeing the line of the practical limits of these techniques~\cite{Dias:2017,Eremets:2019,Loubeyre:2020}.  Complicating the picture, different experiments used different measures to characterize hydrogen's transition to metallicity, from vibrational spectroscopic techniques such as  IR~\cite{Loubeyre:2020}, optical measurements such as reflectance and opacity~\cite{Dias:2017}, and resistivity measurements~\cite{Eremets:2019}, as well as different scales to calibrate pressure. At times, this led to seemingly contradictory results, prompting questions regarding experimental accuracy and reproducibility~\cite{Liu:2017b,Goncharov:2017,Gregoryanz:2020}. \emph{Ab initio} calculations taking into account the quantum fluctuations of the hydrogen nuclei, however, can reconcile some of these differences, finding in the $C2/c$-24 high-pressure phase closing of the electronic gap (and transition to metallicity) before the closing of the optical gap (and transition from transparency to reflectivity)~\cite{Monacelli:2021}.

Additionally, the impractically low $T_c$s of ambient pressure materials are then traded for a much higher $T_c$ in pure metallic hydrogen, but at impractically high pressures! In fact, \emph{ab initio} modeling suggests progressive jumps in $T_c$ with a transition from the molecular (estimated $T_c$ = 356~K near 500~GPa)  to the atomic phase ($T_c$ increasing to 481~K ca.\ 700~GPa), and with an atomic-atomic phase transition at $\sim$1-1.5~TPa driving up $\lambda$ and resulting in an immense estimated $T_c$= 764~K~\cite{McMahon:2011}. To address the hydrogen metallization problem, Ashcroft proposed another strategy -- instead of pure hydrogen, hydrogen-rich metallic alloys could be targeted as putative superconductors~\cite{Ashcroft:2004}. The presence of additional atoms in the hydrogen matrix would confer a \emph{chemical precompression}, thereby lowering the external pressure required to reach the metallic state. Under ambient conditions, the crystal structures adopted by metal hydrides are largely subject to the dictates of balanced oxidation states, hence alkali metal hydrides assume the rock salt structure, and hydrides of +2 metals favor fluorite or Co$_2$Si structures, while trivalent metal hydrides tend towards the BiF$_3$ structure, and so on~\cite{Smithson:2002,Bourgeois:2017}. Much higher hydrogen content would be needed for such hydrogen-rich alloys, as suggested by Ashcroft, and furthermore many of the resulting structures might differ greatly from anything observed at ambient conditions. Defying the recommendations of Matthias, the simultaneous and serendipitous advances in CSP methods meant that theoreticians were well poised to answer this call!

\subsection{Clathrate-like hydrides}\label{subsec12}

They were successful. As described above, calculations on the high-pressure Ca-H system located the CaH$_6$ phase described in Section~\ref{subsec9}~\cite{Wang:2012}, in which Ca atoms are embedded into a hydrogenic sodalite-like clathrate framework. The strong electron-phonon coupling predicted for $Im\bar{3}m$ CaH$_6$ 
can be traced to breathing and rocking modes of the square H$_4$ units of the sodalite framework~\cite{Wang:2012}. The molecular orbital diagram for such an H$_4$ square has, above a filled bonding state, a half-occupied degenerate non-bonding state. Assuming full ionization (integrated charges within atomic basins according to the Quantum Theory of Atoms in Molecules indicate roughly 1 electron per Ca is transferred to the hydrogen network), these orbitals accept the roughly 1/3 electron per H transferred from the Ca atom. This favors symmetry-breaking Jahn-Teller distortions -- key contributions to the electron-phonon coupling -- that lift the degeneracy.

Similar clathrate-like hydrides, as outlined in Section~\ref{subsec9}, were rapidly predicted in several other systems. 
$Im\bar{3}m$ YH$_6$ has been synthesized with a measured $T_c$ of 224~K at 160~GPa~\cite{Troyan:2021a}. Hydrides with even higher hydrogen content have been synthesized as well, including a $P6_3/mmc$ YH$_9$ phase with a $T_c$ of 262~K at ca.\ 180~GPa~\cite{Snider:2021} (a subsequent study reported a slightly lower $T_c$ of 243~K at 200~GPa~\cite{Kong:2021a}), and an isotypic CeH$_9$ phase whose $T_c$ has only been predicted (57~K at 88~GPa and up to 100~K at 130~GPa~\cite{Peng:2017}, or 105-117~K by 200~K~\cite{Salke:2019}). The relatively low pressure required to stabilize CeH$_9$ has been ascribed to strong chemical precompression from the delocalized Ce 4\emph{f} electrons~\cite{Jeon:2020}. The reported $T_c$s for $Fm\bar{3}m$ LaH$_{10}$ (250 and 260~K at 170 and 185~GPa, respectively~\cite{drozdov2019superconductivity,Somayazulu:2019-La}), are in line with theoretical predictions of 257-274~K at 250~GPa~\cite{Liu:2017-La-Y}. Isotypic $Fm\bar{3}m$ YH$_{10}$ is computed to be a  room temperature superconductor with a $T_c$ of 305-327~K at 250~GPa~\cite{Liu:2017-La-Y}. However, YH$_{10}$ has thus far eluded synthetic efforts. Partial doping with lanthanum appears to be one strategy to stabilize YH$_{10}$: a series of ternary (La/Y)H$_{10}$ phases have been experimentally observed, with measured $T_c=$~253~K~\cite{Semenok:2021b}.

Although the $T_c$s of many clathrate-like hydrides are stunning, these phases will surely decompose well above atmospheric pressures! CeH$_9$ is remarkable for the comparatively low, sub-megabar, pressures at which it maintains dynamic stability~\cite{Salke:2019}. In an attempt to preserve the loosely-bound hydrogenic clathrate frameworks that are associated with such strong electron-phonon coupling to lower pressures, one promising strategy involves the addition of a third element in an attempt to further chemically precompress the hydrogenic lattices. The $Fm\bar{3}m$ LaBH$_8$ phase, which is predicted to maintain dynamic stability down to 40~GPa~\cite{diCataldo:2021}, can be derived from LaH$_{10}$ by removing two hydrogen atoms per formula unit and placing  boron atoms into the center of H$_8$ cubes that are empty in LaH$_{10}$~\cite{Liang:2021,Zhang:2022}. LaBH$_8$ has an estimated $T_c$ of 126~K at 50~GPa~\cite{diCataldo:2021} -- and the isostructural LaBeH$_8$ phase is predicted to achieve a $T_c$ of 183~K at 20~GPa~\cite{Zhang:2022}. Other XYH$_8$ phases have been proposed, with a variety of possible elemental combinations ripe for tuning stability and properties~\cite{Zhang:2022}. A second possibility is afforded by the XY$_2$H$_8$ phases that can be constructed by leaving the H$_8$ cubes empty, but stuffing the center of H$_4$ tetrahedra instead. KB$_2$H$_8$ (dynamically stable to 12~GPa~\cite{Gao:2021}) and LaC$_2$H$_8$ (dynamically stable down to 70~GPa~\cite{Durajski:2021}) are two representatives of this structural arrangement, with estimated $T_c$s of 134-146 and 69~K, respectively.

Key to the success of the clathrate-like hydrides is the maintenance of loosely-coordinated networks of hydrogen, rather than condensation into H$_2$ molecules. The effect of H-H interatomic distances on superconductivity can be seen in the MH$_4$ hydrides, which adopt the $I4/mmm$ structure shared with ThCr$_2$Si$_2$~\cite{Shatruk:2019,Bi:2021b}. Hydrogen occupies two inequivalent sites in the ThCr$_2$Si$_2$ structure -- the apical H$_a$ (Wyckoff position 4\emph{e}, Si) and basal H$_b$ (4\emph{d}, Cr). A plethora of metal hydrides have been predicted or synthesized in this structure type under pressure, including Ca~\cite{Wang:2012,Zurek:2018c,Shao:2019b} and Sr~\cite{Hooper:2013b,Wang:2015a}, Sc~\cite{Abe:Sc-2017,Qian:Sc-2017,Zurek:2018b},  Y~\cite{Li:2015a,Liu:2017-La-Y}, and Zr~\cite{Abe:2018-Zr} and rare earths La~\cite{Liu:2017-La-Y}, Ce~\cite{Peng:2017,Li:2019,Salke:2019}, Pr~\cite{Peng:2017,Zhou:2020b,PenaAlvarez:2020}, Pu~\cite{Zhao:2020}, Tb~\cite{Hai:2021}, Eu~\cite{Semenok:2021}, Nd~\cite{Peng:2017,Zhou:2020c}, and Th~\cite{Kvashnin:2018,Semenok:2020c}, making systematic study enticing and useful~\cite{Bi:2021b}. With metal oxidation states ranging from +2 to +4, the formulas of these compounds can be written as M$^{x+}$(H$_b^-$)$_2$(H$_a$)$_2^{(x-2)-}$, with hydridic H$_b$ and a range of charges possible on the H$_a$ atoms. The $T_c$s of these phases are correlated with the length of the H$_a$-H$_a$ contacts, which can behave anywhere from covalently bound H$_2$ units to fully dissociated hydridic anions depending on the metal atom -- similar to the behavior of the X-X bond in ThCr$_2$Si$_2$-type AB$_2$X$_2$ phases~\cite{Hoffmann:1985}. The size of the metal atom can be relevant, as larger atoms will stretch the H$_a$-H$_a$ contact through purely steric interactions, but more important is the valency of the metal atom. Electron transfer from the electropositive metal into the H$_a$-H$_a$ motif directly populates the H$_2$ $\sigma_u^*$ antibonding orbitals, but is also driven by a Kubas-like two-pronged mechanism of H$_2$ $\sigma_g \rightarrow $ M~$d$ donation, and M~$d \rightarrow $ H$_2$ $\sigma_u^*$ back-donation. With enough H$_2$ $\sigma_u^*$ population, the H$_a$ atoms behave in a hydridic fashion lowering the $T_c$, as seen in ZrH$_4$~\cite{Abe:2018-Zr} and ThH$_4$~\cite{Kvashnin:2018}. Donation of sufficient electron density to weaken, but not fully break, the H$_a$-H$_a$ bonding interaction results in a much higher DOS at $E_F$ and enhanced $T_c$, as in YH$_4$~\cite{Li:2015a}.

\subsection{Covalent hydrides}\label{subsec13}

\begin{figure}[!h]
\begin{center}
\includegraphics[width=\figurewidth]{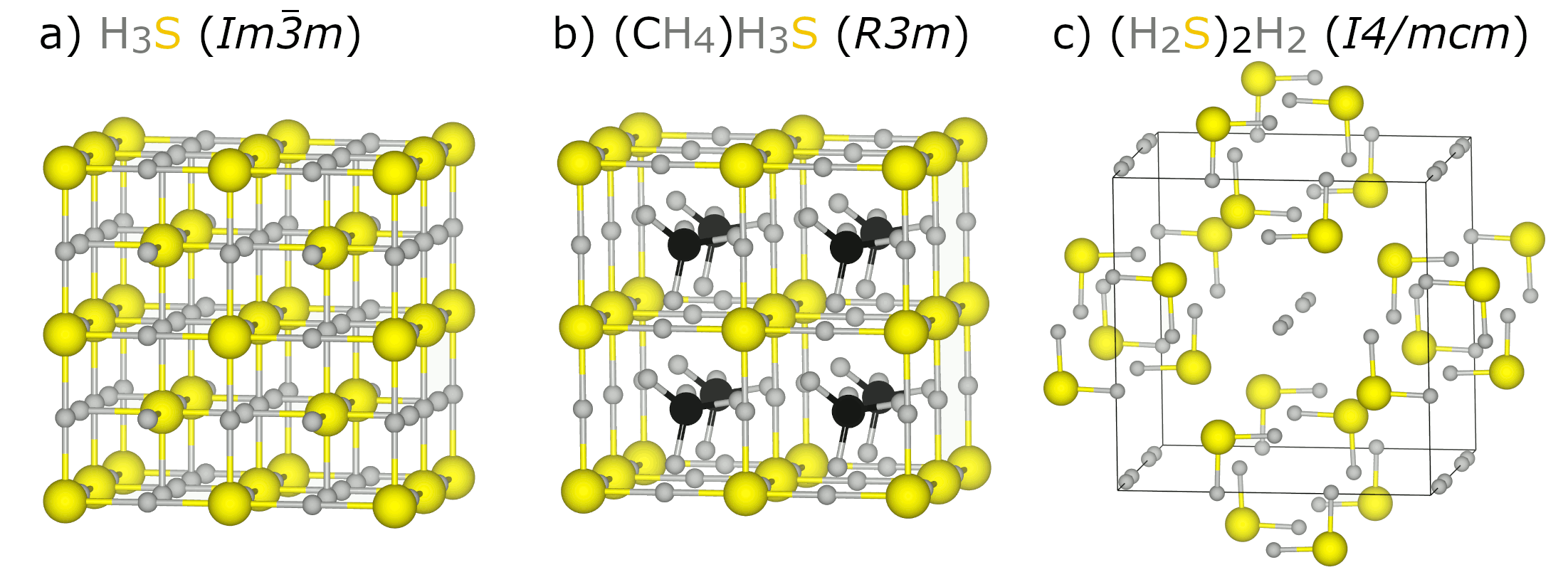}
\end{center}
\caption{The covalent hydride $Im\bar{3}m$ H$_3$S~\cite{Drozdov:2015a} (a) represents a breakthrough in high-$T_c$ conventional superconductivity. Peaks in the electronic DOS near $E_F$ have prompted numerous investigations on doped versions of H$_3$S, discovering phases such as the methane-intercalated (CH$_4$)H$_3$S = CSH$_7$~\cite{Cui:2020} (b), among many others. A recent synthesis of a room-temperature superconductor consisting of carbon, sulfur, and hydrogen generated even more momentum, with diffraction analysis performed at pressures below those where the room-temperature superconducting transition were observed, revealing the presence of a phase based on the $I4/mcm$ Al$_2$Cu-like structure adopted by the van der Waals (H$_2$S)H$_2$ phase~\cite{Strobel:2011,Lamichhane:2021,Goncharov:2022} (c).  
\label{fig:H3S}}
\end{figure}

The first hydride to top the charts, as it were, was not of the metal clathrate-like family, but instead came from attempts to metallize H$_2$S. Theory identified an H$_2$S compound that was computed to possess a $T_c$ of 80~K at 160~GPa~\cite{Li:2014}. Experimental confirmation followed shortly thereafter, finding a phase with $T_c$ $<$ 100~K, but in the process a higher-temperature preparation method yielded a sample with a $T_c$ of 203~K at 150~GPa~\cite{Drozdov:2015a}. A few years before, synthetic exploration into the (H$_2$S)$_2$H$_2$ stoichiometry found a phase stabilized by pressure-induced hydrogen bonding above 3.5~GPa~\cite{Strobel:2011}. This inspired CSP investigations of the H$_3$S stoichiometry, which found an $R{3}m$ phase with $T_c=$~155-166~K at 130~GPa~\cite{Duan:2014}. By 180~GPa this structure transitioned to one with $Im\bar{3}m$ symmetry (Figure~\ref{fig:H3S}a) for which the estimated $T_c$ was 191-204~K at 200~GPa. Serendipitously the experimental~\cite{Drozdov:2015a} and theoretical~\cite{Duan:2014} manuscripts appeared at nearly the same time. Subsequent XRD studies supported the identification of the 203~K superconductor as $Im\bar{3}m$ H$_3$S~\cite{Einaga:2016}, though other structures have been proposed~\cite{Li:2016,Gordon:2016,Ishikawa:2016,Akashi:2016,Yao:2018,Laniel:2020}.

A host of studies on the H$_3$S superconductor have followed, exploring the isotope effect, role of anharmonicity, and possible quantum effects~\cite{Errea:2016,Ortenzi:2016,Durajski:2015,Durajski:2016,Gorkov:2016,Goncharov:2016,Jarlborg:2016,Bussmann-Holder:2017,Szczesniak:2017,Azadi:2017,Arita:2017}. The inclusion of quantum nuclear motions lowered the pressure where the less symmetric $R3m$ phase was predicted to transition to the  $Im\bar{3}m$ structure with symmetric H-S bonds into the range of pressures where high $T_c$s had been measured~\cite{Errea:2016}. One of the most striking features of the electronic structure of H$_3$S are a pair of van Hove singularities bracketing $E_F$~\cite{Quan:2016,Sano:2016}. Shifting the position of $E_F$, potentially by doping, could increase the number of states that can participate in the electron-phonon coupling mechanism, and therefore increase the $T_c$ of the system.

Doping is a common strategy used for precise tuning of $E_F$, and computations using the virtual crystal approximation (VCA) suggested that the addition of a little bit of phosphorus, carbon, or silicon could raise the $T_c$ into the room-temperature regime $>$ 280~K~\cite{Heil:2015,Ge:2016,Ge:2020}. In the VCA, alchemical pseudoatoms are constructed from weighted averages of the component atom potentials. The resulting chemical chimeras, however, cannot accurately model the local structural and electronic effects that arise when one atom is replaced with an entirely different element. This throws off, in particular, the very dynamical response properties that one must calculate carefully to obtain reasonable  estimates of  $T_c$. Additional studies based on actual doped H$_3$S models constructed as supercells have sought to explore the local effects of doping~\cite{Liu:2018c,Amsler:2019,Guan:2021,Wang:2022}, although the calculation of dynamical properties of the requisite large unit cells can be prohibitively expensive. 

One particularly promising system involved the addition of carbon to the H$_3$S lattice by way of methane intercalation~\cite{Cui:2020,Sun:2020, Harshman:2022}. Stoichiometries that are a linear combination of CH$_4$ and H$_3$S (the most simple of which is CSH$_7$) proved especially interesting. They yielded a variety of dynamically stable (although energetically metastable) structures, which differed in the orientation of the methane molecules encapsulated in the H$_3$S lattice. Some of the highest $T_c$s predicted for these phases were 181~K at 100~GPa for $I\bar{4}3m$~\cite{Sun:2020}, and 181-194~K for the $R3m$ symmetry structure~\cite{Cui:2020} shown in Figure~\ref{fig:H3S}b. 

Independently, photochemical synthesis in the C-S-H system yielded the first report of room-temperature superconductivity, achieving a $T_c$ of 288~K at 267~GPa~\cite{Snider:2020}. This report has inspired a slew of follow-up work and much debate~\cite{Hirsch:2021,Gubler:2022,Goncharov:2022}. XRD analysis performed at pressures below the purported room-temperature superconducting transition~\cite{Lamichhane:2021,Bykova:2021} is consistent with the Al$_2$Cu geometry (as well as an orthorhombic $Pnma$ variant) associated with CH$_4$-H$_2$~\cite{Somayazulu:1996} and H$_2$S-H$_2$~\cite{Strobel:2011} (Figure~\ref{fig:H3S}c). This may suggest an overall stoichiometry of [(CH$_4$)$_2$H$_2$]$_x$[(H$_2$S)$_2$H$_2$]$_y$ for the room-temperature C-S-H superconductor, although subsequent phase transitions at higher pressure to the high-$T_c$ superconducting phase cannot be ruled out.
In fact, additional studies indicate just such a structural transition  occurs to form the room-temperature superconducting phase, with indications of methane signatures in the Raman spectra~\cite{Goncharov:2022}. As was the case for the binary H$_3$S system, it appears that a panoply of metastable phases may be accessible by slight variations on synthetic procedure, in particular on carbon content~\cite{Smith:2021}, offering plenty of space for further experimental and theoretical discoveries.

In addition to sulfur-based covalent hydrides, phosphorus hydrides have sparked interest after compression of a phosphine (PH$_3$) sample yielded a material that became superconducting at 30~K at 83~GPa, increasing to 103~K at 207~GPa~\cite{Drozdov:2015}. The structure and composition of the responsible phase or phases was unclear, prompting an array of follow-up studies levying CSP techniques to identify plausible compounds~\cite{Fu:2016,Liu:2016,Shamp:2016,Flores-Livas:2016,Bi:2017}. Pressure was found to drive the decomposition of phosphine into a variety of products with stoichiometries including PH, PH$_2$, PH$_3$, and more. A predicted $C2/m$ PH$_3$ phase~\cite{Liu:2016} featuring P-P bonds (in contrast to the H$_3$S superconductor, which has no S-S bonding) was estimated to be superconducting below 83~K at 200~GPa, in line with the experimental values. Another study suggested that multiple metastable decomposition products of phosphine, including those with PH and PH$_2$ stoichiometries, might in combination be responsible for the observed superconductivity~\cite{Flores-Livas:2016}. PH$_2$ phases with $C2/m$ and $I4/mmm$ symmetries, differing by a tilt in the component H-P-H moieties, had estimated $T_c$s of 76 and 70~K, respectively~\cite{Shamp:2016}. Later, another set of PH$_2$ phases were proposed consisting of simple cubic layers of phosphorus capped with hydrogen atoms and further intercalated with H$_2$ molecules acting as Coulombic spacing agents~\cite{Bi:2017}. At 80~GPa, these structures had estimated $T_c$s ca. 30~K, similar to the values that were measured. Raman spectroscopic measurements provided evidence for phosphine dimerization coupled with dehydrogenation under pressure, yielding compositions such as P$_2$H$_4$ and P$_3$H$_6$~\cite{Liu:2018b,Yuan:2019}. In these phases, low temperatures were required to maintain stability at multimegabar pressures. 
The $T_c$ of a predicted $C2/m$ P$_4$H$_6$ phase was estimated to be 67~K at 200~GPa~\cite{Yuan:2019}.

The plethora of metastable P-H compounds under pressure has prompted computational investigations into ternary systems containing phosphorus and hydrogen. Above 250~GPa, an $R\bar{3}$ LiP$_2$H$_{14}$ phase, consisting of P@H$_9$ clusters spaced by Li atoms as well as isolated H atoms, achieves an estimated $T_c$ of 169~K at 230~GPa (where it is metastable)~\cite{Li:2020}. $Pm\bar{3}$ LiPH$_6$, a colored variant of the A15 crystal structure adopted by intermetallic superconductors Nb$_3$Ge~\cite{Gavaler:1973} and Nb$_3$Sn~\cite{Matthias:1954}, has an estimated $T_c$ of 150-167~K at 200~GPa (where it is metastable)~\cite{Shao:2019}. In the S-P-H system, obviously tantalizing for its connection to the H$_3$S superconductor as well as to phosphine derivatives, relatively low $T_c$s were predicted for phases on the high-pressure convex hulls, but low-lying metastable structures based on phosphorus substitution into the $Im\bar{3}m$ H$_3$S lattice were promising, including $Im\bar{3}m$ S$_7$PH$_{24}$, which had an estimated $T_c$ of 183~K at 200~GPa~\cite{Geng:2022}.

\section{Conclusion}\label{sec6}
Although the entirety of the lived human experience resides within a vastly narrow pressure range, the universe is not so simple. The chemistry we know at 1~atmosphere is not the chemistry of Jupiter, Saturn, or even the center of our own planet Earth. Starting from the periodic table itself, the ramifications of pressure are rapidly found to alter elemental behavior and, consequently, how the elements interact with one another to form new and bizarre phases. Potassium, in its guise as a ``transition metal", enjoys all manner of new chemical interactions -- in compound formation and in the wildly complex electride elemental structures it adopts. Cesium can become anionic, and helium takes an active role in stabilizing a network of sodium and interstitial quasiatoms. Strange geometrical and bonding motifs from clusters to networks abound.

Yet not only are the structures of phases -- electronic and crystalline -- molded by high pressure, but high-pressure studies have revolutionized the search for high-temperature superconductivity. Pressure induces superconductivity in a plethora of elements, and drives the formation of phases containing structural motifs whose atomic vibrations can be strongly coupled to the underlying electronic structure. From the clathrate-like LaH$_{10}$ to the covalent H$_3$S -- and the intensely-discussed CSH room-temperature superconductor -- the playing field of high-pressure materials is a promising one for the future. 

\section*{Acknowledgments}

We acknowledge the NSF (DMR-1827815) for financial support.  This material is based upon work supported by the U.S.\ Department of Energy, Office of Science, Fusion Energy Sciences under Award Number DE-SC0020340 to E.Z. K.P.H.\ thanks the US Department of Energy, National Nuclear Security Administration, through the Chicago-DOE Alliance Center under Cooperative Agreement Grant No.\ DE-NA0003975 for financial support. We thank Giacomo Scilla for his help in editing and preparing the manuscript.

\backmatter

\bibliography{bibliography.bib}


\end{document}